 \documentclass[twocolumn]{aastex631}
\newcommand{\thetasep}{\theta_{\rm sep}}

\shorttitle{Follow-up survey for the BBH merger GW200224\_222234 using Subaru/HSC and GTC/OSIRIS}
\shortauthors{T. Ohgami, J. Becerra Gonz\'alez and N. Tominaga et al.}
\graphicspath{{./}{figures/}}

\begin{document}

\title{Follow-up survey for the binary black hole merger GW200224\_222234 using Subaru/HSC and GTC/OSIRIS}

\correspondingauthor{Takayuki Ohgami}
\email{takayuki.ohgami@nao.ac.jp}
\correspondingauthor{Josefa Becerra Gonz\'alez}
\email{jbecerra@iac.es}
\correspondingauthor{Nozomu Tominaga}
\email{nozomu.tominaga@nao.ac.jp}

\author[0000-0001-6755-8285]{Takayuki Ohgami} 
\affiliation{Division of Science, National Astronomical Observatory of Japan, 2-21-1, Osawa, Mitaka, Tokyo 181-8588, Japan}

\author[0000-0002-6729-9022]{Josefa Becerra Gonz\'alez} 
\affiliation{Instituto de Astrof\'isica de Canarias (IAC), E-38200 La Laguna, Tenerife, Spain}
\affiliation{Universidad de La Laguna (ULL), Departamento de Astrof\'isica, E-38206 La Laguna, Tenerife, Spain}

\author[0000-0001-8537-3153]{Nozomu Tominaga} 
\affiliation{Division of Science, National Astronomical Observatory of Japan, 2-21-1, Osawa, Mitaka, Tokyo 181-8588, Japan}
\affiliation{Department of Astronomical Science, School of Physical Sciences, The Graduate University of Advanced Studies, (SOKENDAI), 2-21-1 Osawa, Mitaka, Tokyo 181-8588, Japan}
\affiliation{Kavli Institute for the Physics and Mathematics of the Universe (WPI), The University of Tokyo, 5-1-5 Kashiwanoha, Kashiwa, Chiba 277-8583, Japan}
\affiliation{Department of Physics, Faculty of Science and Engineering, Konan University, 8-9-1 Okamoto, Kobe, Hyogo 658-8501, Japan}

\author[0000-0001-7449-4814]{Tomoki Morokuma} 
\affiliation{Planetary Exploration Research Center, Chiba Institute of Technology, 2-17-1 Tsudanuma, Narashino, Chiba 275-0016, Japan}
\affiliation{Institute of Astronomy, Graduate School of Science, The University of Tokyo, 2-21-1 Osawa, Mitaka, Tokyo 181-0015, Japan}
\affiliation{Kavli Institute for the Physics and Mathematics of the Universe (WPI), The University of Tokyo, 5-1-5 Kashiwanoha, Kashiwa, Chiba 277-8583, Japan}

\author[0000-0001-6161-8988]{Yousuke Utsumi} 
\affiliation{Kavli Institute for Particle Astrophysics and Cosmology (KIPAC), SLAC National Accelerator Laboratory, Stanford University, 2575 Sand Hill Road, Menlo Park, CA 94025, USA}

\author[0000-0001-5322-5076]{Yuu Niino} 
\affiliation{Institute of Astronomy, Graduate School of Science, The University of Tokyo, 2-21-1 Osawa, Mitaka, Tokyo 181-0015, Japan}
\affiliation{Research Center for the Early Universe, Graduate School of Science, The University of Tokyo, 7-3-1 Hongo, Bunkyo-ku, Tokyo 113-0033, Japan}

\author[0000-0001-8253-6850]{Masaomi Tanaka} 
\affiliation{Astronomical Institute, Tohoku University, Sendai, Miyagi 980-8578, Japan}

\author[0000-0001-6595-2238]{Smaranika Banerjee} 
\affiliation{Astronomical Institute, Tohoku University, Sendai, Miyagi 980-8578, Japan}

\author[0000-0002-5391-5568]{Fr\'ed\'erick Poidevin} 
\affiliation{Instituto de Astrof\'isica de Canarias (IAC), E-38200 La Laguna, Tenerife, Spain}
\affiliation{Universidad de La Laguna (ULL), Departamento de Astrof\'isica, E-38206 La Laguna, Tenerife, Spain}

\author[0000-0002-0433-9656]{Jose Antonio Acosta-Pulido} 
\affiliation{Instituto de Astrof\'isica de Canarias (IAC), E-38200 La Laguna, Tenerife, Spain}
\affiliation{Universidad de La Laguna (ULL), Departamento de Astrof\'isica, E-38206 La Laguna, Tenerife, Spain}

\author[0000-0002-2807-6459]{Ismael P\'erez-Fournon} 
\affiliation{Instituto de Astrof\'isica de Canarias (IAC), E-38200 La Laguna, Tenerife, Spain}
\affiliation{Universidad de La Laguna (ULL), Departamento de Astrof\'isica, E-38206 La Laguna, Tenerife, Spain}

\author[0000-0002-3348-4035]{Teo Mu\~noz-Darias} 
\affiliation{Instituto de Astrof\'isica de Canarias (IAC), E-38200 La Laguna, Tenerife, Spain}
\affiliation{Universidad de La Laguna (ULL), Departamento de Astrof\'isica, E-38206 La Laguna, Tenerife, Spain}

\author[0000-0001-6156-238X]{Hiroshi Akitaya} 
\affiliation{Planetary Exploration Research Center, Chiba Institute of Technology, 2-17-1 Tsudanuma, Narashino, Chiba 275-0016, Japan}
\affiliation{Hiroshima Astrophysical Science Center, Hiroshima University, 1-3-1 Kagamiyama, Higashi-Hiroshima, Hiroshima 739-8526, Japan}

\author[0000-0001-7592-9285]{Kenshi Yanagisawa} 
\affiliation{Subaru Telescope, National Astronomical Observatory of Japan, 650 North A'ohoku Place, Hilo, HI 96720, USA}
\affiliation{Hiroshima Astrophysical Science Center, Hiroshima University, 1-3-1 Kagamiyama, Higashi-Hiroshima,
Hiroshima 739-8526, Japan}

\author[0000-0001-5946-9960]{Mahito Sasada} 
\affiliation{Hiroshima Astrophysical Science Center, Hiroshima University, 1-3-1 Kagamiyama, Higashi-Hiroshima,
Hiroshima 739-8526, Japan}
\affiliation{Mizusawa VLBI Observatory, National Astronomical Observatory of Japan, 2-12 Hoshigaoka, Mizusawa,
Oshu, Iwate 023-0861, Japan}

\author[0000-0002-9948-1646]{Michitoshi Yoshida} 
\affiliation{Subaru Telescope, National Astronomical Observatory of Japan, 650 North A'ohoku Place, Hilo, HI 96720, USA}

\author[0000-0002-5652-6525]{Mirko Simunovic} 
\affiliation{Subaru Telescope, National Astronomical Observatory of Japan, 650 North A'ohoku Place, Hilo, HI 96720, USA}

\author[0000-0001-5797-6010]{Ryou Ohsawa} 
\affiliation{Institute of Astronomy, Graduate School of Science, The University of Tokyo, 2-21-1 Osawa, Mitaka, Tokyo 181-0015, Japan}
\affiliation{JASMINE Project, National Astronomical Observatory of Japan, 2-21-1, Osawa, Mitaka, Tokyo 181-8588, Japan}

\author{Ichi Tanaka} 
\affiliation{Subaru Telescope, National Astronomical Observatory of Japan, 650 North A'ohoku Place, Hilo, HI 96720, USA}

\author[0000-0003-4143-4246]{Tsuyoshi Terai} 
\affiliation{Subaru Telescope, National Astronomical Observatory of Japan, 650 North A'ohoku Place, Hilo, HI 96720, USA}

\author[0000-0002-9104-8979]{Yuhei Takagi} 
\affiliation{Subaru Telescope, National Astronomical Observatory of Japan, 650 North A'ohoku Place, Hilo, HI 96720, USA}

\author{The J-GEM collaboration}



\begin{abstract} 

The LIGO/Virgo detected a gravitational wave (GW) event, named GW200224\_222234 (a.k.a. S200224ca) and
classified as a binary-black-hole coalescence, on February 24, 2020.
Given its relatively small localization skymap (71\,deg$^2$ for a 90\% credible region; revised to 50\,deg$^2$ in GWTC-3), we performed target-of-opportunity observations using the Subaru/Hyper Suprime-Cam (HSC) in the $r2$- and $z$-bands.
Observations were conducted on February 25 and 28 and March 23, 2020, with the first epoch beginning 12.3\,h after the GW detection.
The survey covered the highest probability sky area of 56.6\,deg$^2$, corresponding to a 91\% probability.
This was the first deep follow-up ($m_{r}\gtrsim24,~m_{z}\gtrsim23$) for a binary-black-hole merger covering $>$90\% of the localization.
By performing image subtraction and candidate screening including light
 curve fitting with transient templates and examples, we found 22 off-nucleus transients that were not ruled out as the counterparts of GW200224\_222234 with only our Subaru/HSC data.
We also performed GTC/OSIRIS spectroscopy of the probable host galaxies for five candidates; two are likely to be located within the 3D skymap, whereas the others are not.
In conclusion, 19 transients remain as possible optical counterparts of GW200224\_222234; however, we could not identify a unique promising counterpart.
If there are no counterparts in the remaining candidates, the upper limits of optical luminosity are $\nu L_{\nu} < 5.2^{+2.4}_{-1.9}\times 10^{41}$ erg~s$^{-1}$ and $\nu L_{\nu} < 1.8^{+0.8}_{-0.6}\times 10^{42}$ erg~s$^{-1}$ in the $r2$- and $z$-bands, respectively, at $\sim$12\,h after GW detection. We also discuss improvements in the strategies of optical follow-ups for future GW events.
\end{abstract}

\keywords{Gravitational waves (678); Black holes (162); Surveys (1671)}


\section{Introduction} \label{sec:intro}

In general relativity, massive objects radiate energy via the distortion of space time when their motion is accelerated.
This energy radiation predicted by Einstein is called a gravitational wave \citep[GW;][]{1916SPAW.......688E, 1918SPAW.......154E}.
Astronomical objects or phenomena are expected to be sources of GW signals with a large amplitude, which may be detected by current instruments.
For example, binary systems composed of compact objects such as black holes (BHs) or neutron stars (NSs) emit strong GWs at their coalescence.

In the second observation run of the GW interferometers LIGO and Virgo, they detected a GW signal from a binary-NS (BNS) coalescence using three detectors, and the localization area was constrained to 28\,deg$^{2}$ for a 90\% credible region \citep[GW170817;][]{2017ApJ...848L..12A}.
The electromagnetic (EM) counterpart of GW events was observed for the first time by multiple observatories across the EM spectrum from radio to $\gamma$-rays \citep{2017PASA...34...69A, 2017ApJ...848L..33A, 2017ApJ...848L..19C, 2017Sci...358.1556C, 2017ApJ...848L..17C, 2017ApJ...848L..29D, 2017Sci...358.1570D, 2017Sci...358.1565E, 2017Sci...358.1559K, 2017ApJ...850L...1L, 2017ApJ...848L..18N, 2017Natur.551...67P, 2017Natur.551...75S, 2017ApJ...848L..16S, 2017Natur.551...71T, 2017ApJ...848L..27T, 2018PASJ...70...28T, 2018PASJ...70....1U, 2017ApJ...848L..24V}.
It was demonstrated that BNS mergers are accompanied by explosive EM emissions called kilonova \citep[e.g.,][]{2013ApJ...774...25K, 2017PhRvD..96l3012S, 2017PASJ...69..102T, 2018ApJ...865L..21K, 2017ApJ...850L..37P, 2018A&A...615A.132R,2020ApJ...901...29B}.

In contrast to NS mergers, binary-BH (BBH) mergers are not considered to be accompanied by any EM emission.
However, the Fermi Gamma-ray Burst Monitor (GBM) reported the presence of a weak $\gamma$-ray transient after the detection of GW150914 \citep{2016ApJ...826L...6C}.
Although the physical association between GWs and $\gamma$-ray signals
is ambiguous, and it is unclear whether the $\gamma$-ray signal is truly astronomical because of the low flux, various scenarios have been proposed, such as BBHs surrounded by pre-existing material, for example, a circumbinary disc (\citealt{2018MNRAS.480.4732M}), an accretion disc surrounding a galactic center BH (\citealt{2017ApJ...835..165B}), and remnants of gravitational collapse (\citealt{2017NewA...51....7J}).
Another case is the detection of an optical counterpart candidate of GW190521 by the Zwicky Transient Facility (ZTF19abanrhr).
\citet{2020PhRvL.124y1102G} suggested that the EM flare is consistent with the behavior expected from a kicked BBH merger in an accretion disk of an active galactic nucleus (AGN, \citealt{2019ApJ...884L..50M}) and ruled out other scenarios (for example, the intrinsic variability of AGN, supernova, microlens, tidal disruption). \cite{2022arXiv220913004G} comprehensively searched for EM counterparts to BBH mergers and identified nine candidates.
To test various scenarios for EM radiation from a BBH merger, it is still important to perform follow-ups of the BBH mergers.

The LIGO/Virgo collaboration began their third observation run (O3) in April 2019 and detected a GW event named GW200224\_222234 (a.k.a. S200224ca) using three detectors on February 24, 2020 at 22:22 UTC \citep{2021arXiv211103606T, 2020GCN.27184....1L}.
They released a preliminary localization skymap derived using software called BAYESTAR \citep{2016PhRvD..93b4013S} on February 24, 2020 at 22:32 UTC.
In this release, the GW event was classified as a BBH coalescence with a $>99$\% confidence level and a false alarm rate of $1.6\times 10^{-11}$\,Hz (approximately one in 1975~years).
The luminosity distance was $1583\pm331$\,Mpc, corresponding to a redshift of $0.30\pm0.05$, and the 90\% localization sky area was as narrow as 71\,deg$^2$.

Upon receiving the alert for GW200224\_222234, the Japanese collaboration for Gravitational wave ElectroMagnetic follow-up \citep[J-GEM;][]{2016PASJ...68L...9M, 2018PASJ...70....1U, 2021PTEP.2021eA104S} triggered a target-of-opportunity (ToO) observation to search for its EM counterpart on February 25, 2020 at 10:43 UTC \citep{2020GCN.27205....1O} using Hyper Suprime-Cam \citep[HSC;][]{2018PASJ...70S...3F, 2018PASJ...70...66K, 2018PASJ...70S...2K, 2012SPIE.8446E..0ZM, 2018PASJ...70S...1M}, which is a wide-field imager installed on the prime focus of the Subaru Telescope.
Its field of view (FoV) of 1.77\,deg$^2$ is the largest among current 8\,m-class telescopes, which makes the Subaru/HSC the most efficient instrument for the optical survey.
The first exposure commenced at approximately 12.3\,h after GW detection, and the observation area reached 56.6\,deg$^2$.
We also conducted additional ToO observations using the Subaru/HSC on February 28 and March 23, 2020, in the same fields as the first epoch observation.
The LIGO and Virgo collaboration published the GWTC-3 catalog \citep{2021arXiv211103606T} including the observations O1, O2, O3a, and O3b and also released the reanalyzed the localization skymap.
The 90\% localization sky area was updated from 71\,deg$^2$ to 50 deg$^2$.
Our observation area corresponds to a cumulative probability of 91\% in the updated skymap.

In this paper, we describe the details of the follow-ups of GW200224\_222234 using the Subaru/HSC, the candidate selection algorithm, and a list of candidates, including spectroscopic observations of the probable host galaxies of the five candidates to measure their spectroscopic redshifts using the Optical System for Imaging and low-Intermediate-Resolution Integrated Spectroscopy\footnote{\url{http://www.gtc.iac.es/instruments/osiris/}} (OSIRIS), which is an imager and spectrograph for the optical wavelength range, installed in the 10.4-\,m Gran Telescopio CANARIAS (GTC).
All magnitudes are given in AB magnitudes.

\section{Observations with the Subaru/HSC and data analysis} \label{sec:HSC observation}
\subsection{ToO observation} \label{ssec:ToO}
\begin{figure*}[ht!]
    \begin{center}
        \includegraphics[width=.9\hsize]{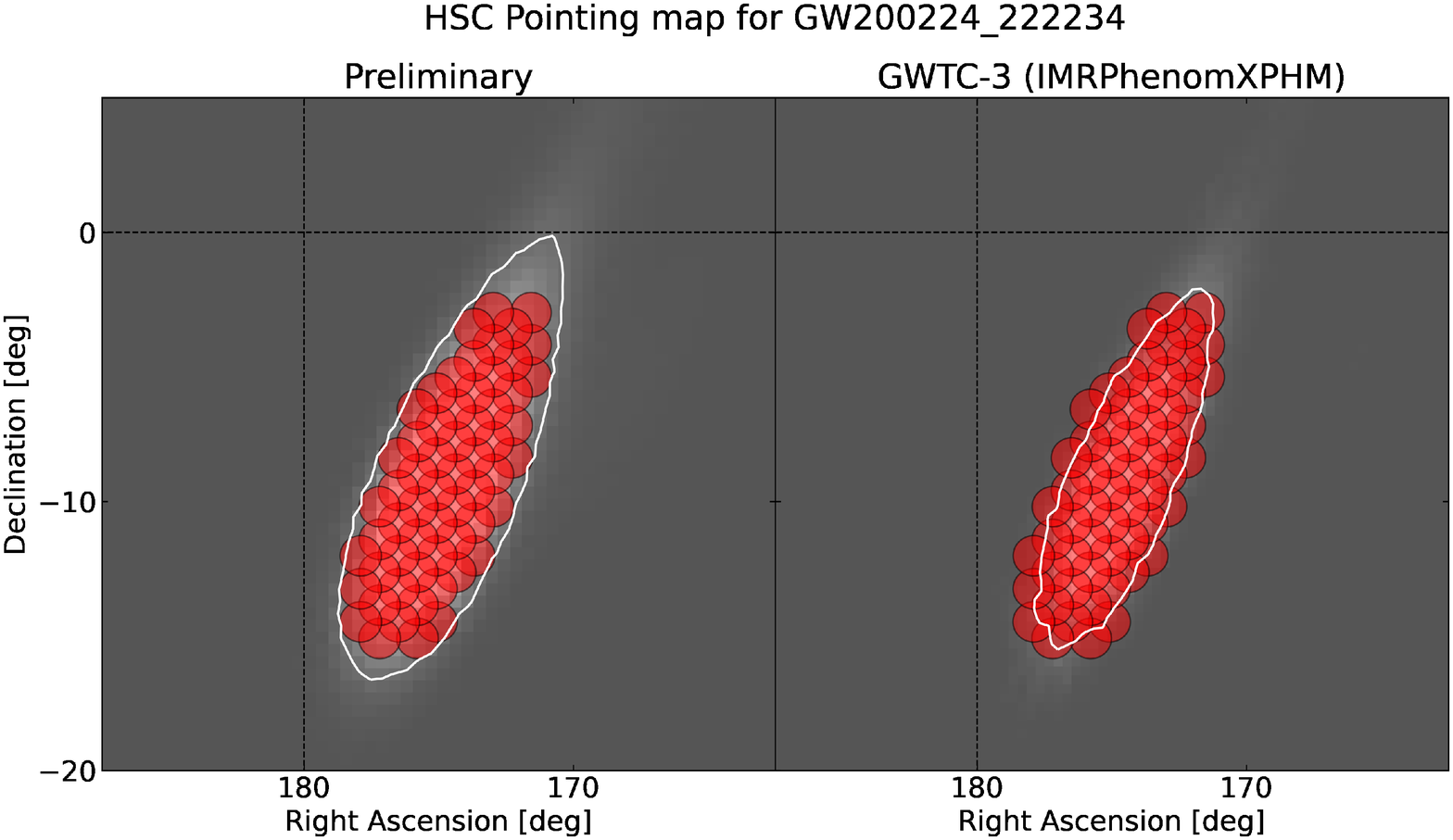}
    \end{center}
    \caption{Observation pointings with the Subaru/HSC (red filled circles). The left and right panels are the preliminary localization skymap of GW200224\_222234 \citep[BAYESTAR;][]{2020GCN.27184....1L} and a refined version using the IMRPhenomXPHM model \citep[GWTC-3 catalog;][]{2021arXiv211103606T}, respectively. The white contour lines indicate the 90\% credible regions.}
    \label{fig:pointings}
\end{figure*}

We conducted optical imaging observations using the Subaru/HSC on February 25 (Day~1), 28 (Day~4), and March 23 (Day~28), 2020, in the $r2$- and $z$-bands.
To trigger our ToO follow-up as rapidly as possible, we conducted the first observations in the $r2$-band, which had already been set in the Subaru/HSC at the time of the GW alert.
The first exposure commenced on February 25, 2020 at 10:43 UTC, corresponding to 12.3\,h after GW detection.
We selected 60 observation pointings to cover the high probability area in the BAYSTAR skymap for the HEALPix grid with a resolution of \texttt{NSIDE}~$=64$, which corresponds to 0.84\,deg$^2$\,pixel$^{-2}$, enabling overlap of the FoVs.
Figure \ref{fig:pointings} shows the survey pointing map.
Our survey area covered 56.6\,deg$^2$, corresponding to a cumulative probability of 91\% in the localization skymap refined in the GWTC-3 catalog using the IMRPhenomXPHM model \citep{2021PhRvD.103j4056P}, which models higher-order spherical harmonics and spin precession.
In this study, we used the localization skymap created using the IMRPhenomXPHM model as the 3D localization of GW200224\_222234.

On February 25 and 28, 2020, we observed these pointings with 30\,s of exposure each in pointing ID order and revisited them in each band.
The revisits were conducted at least 1\,h apart with a 1$'$ offset in each pointing to fill the CCD gaps.
Note that some areas were observed again at a time interval of less than 1 h because of the overlap between each pointing.
On March 23, 2020, we changed the exposure time and the order of the pointings.
The pointings at higher elevation were observed earlier.
In the $z$-band, we took the shots with a 35-\,s exposure each and revisited them.
In the $r2$-band, we took the shots with 50-\,s exposures for the first 32 pointings and 70-\,s exposures for the remaining 28 pointings.
The central coordinates and exposure times are shown in Table \ref{tab:pointings} (This table is published in its entirety in the machine-readable format).

\begin{deluxetable*}{lcccccccc}
\tablecaption{Central coordinates and exposure times of the observation pointings with the Subaru/HSC. \label{tab:pointings}}
\tablewidth{0pt}
\tablehead{
\colhead{} &
\multicolumn{2}{c}{Central Coordinate} &
\multicolumn{6}{c}{Exposure Time} \\
\colhead{Pointing ID} &
\colhead{R.A. (J2000.0)} &
\colhead{Decl. (J2000.0)} &
\multicolumn{2}{c}{2020-02-25 (Day~1)} &
\multicolumn{2}{c}{2020-02-28 (Day~4)} &
\multicolumn{2}{c}{2020-03-23 (Day~28)} \\
\colhead{} &
\colhead{(HH:MM:SS.ss)} &
\colhead{(DD:MM:SS.s)} &
\colhead{$r2$} &
\colhead{$z$} &
\colhead{$r2$} &
\colhead{$z$} &
\colhead{$r2$} &
\colhead{$z$}
}
\startdata
00 & 11:40:18.75 & $-$14:28:39.0 & 30\,s$\times2$ & 30\,s$\times2$ & 30\,s$\times2$ & 30\,s$\times2$ & 50\,s & 35\,s$\times2$ \\
01 & 11:34:41.25 & $-$12:01:28.9 & 30\,s$\times2$ & 30\,s$\times2$ & 30\,s$\times2$ & 30\,s$\times2$ & 70\,s & 35\,s$\times2$ \\
02 & 11:31:52.50 & $-$10:11:59.7 & 30\,s$\times2$ & 30\,s$\times2$ & 30\,s$\times2$ & 30\,s$\times2$ & 50\,s & 35\,s$\times2$ \\
03 & 11:43:07.50 & $-$15:05:41.2 & 30\,s$\times2$ & 30\,s$\times2$ & 30\,s$\times2$ & 30\,s$\times2$ & 70\,s & 35\,s$\times2$ \\
04 & 11:37:30.00 & $-$12:38:08.3 & 30\,s$\times2$ & 30\,s$\times2$ & 30\,s$\times2$ & 30\,s$\times2$ & 70\,s & 35\,s$\times2$ \\
$\vdots$ & $\vdots$ & $\vdots$ & $\vdots$ & $\vdots$ & $\vdots$ & $\vdots$ & $\vdots$ & $\vdots$ \\
\enddata
\tablecomments{This table is published in its entirety in the machine-readable format. A portion is shown here for guidance regarding its form and content.}
\end{deluxetable*}

\subsection{Data reduction and image subtraction} \label{ssec:data reduction}
We reduced the observational data using hscPipe v4.0.5 \citep{2018PASJ...70S...5B}, which is a standard analysis pipeline for the HSC.
This pipeline provides full packages for data analyses, which include image subtraction and source detection.
We evaluated the limiting magnitudes of the stacked images in each epoch as follows:
We assumed an aperture with a diameter of twice the full width at half
maximum (FWHM) of the point spread function (PSF) and distributed it
randomly, avoiding existing sources.
Measuring the $5\sigma$ deviations of fluxes in this aperture, we obtained a map of the $5\sigma$ limiting magnitudes for each stack image.
Table \ref{tab:limmag} lists the mode\footnote{The mode values are the peak values of the histograms of the $5\sigma$ limiting magnitudes. Here, we distributed the $5\sigma$ limiting magnitudes into ten bins ranging between the maximum and minimum values.}, maximum, and minimum values of the $5\sigma$ limiting magnitudes.
The $5\sigma$ limiting magnitudes scattered with a range of $\sim1$\,mag, and there were differences of approximately 1\,mag between those of the $r2$- and $z$-bands.
Note that the limiting magnitudes in the $r2$-band on March 23, 2020, had a small positional dependence owing to the different exposure time.

\begin{table}[ht!]
    \caption{Mode, maximum, and minimum values of the $5\sigma$ limiting magnitudes in the stacked images.}
    \centering
    \begin{tabular}{lcccc}
        \hline \hline
        Date & Filter & \multicolumn{3}{c}{Limiting magnitude (AB)} \\
        (Days from GW detection)  & & mode & max & min \\
        \hline
        2020-02-25 & $r2$ & 25.29 & 25.71 & 24.51 \\
        (Day 1)    & $z$  & 23.56 & 24.14 & 22.84 \\
        2020-02-28 & $r2$ & 24.61 & 25.13 & 23.98 \\
        (Day 4)    & $z$  & 23.21 & 23.76 & 22.54 \\
        2020-03-23 & $r2$ & 25.53 & 26.25 & 24.64 \\
        (Day 28)   & $z$  & 23.83 & 24.64 & 22.85 \\
        \hline
    \end{tabular}
    \label{tab:limmag}
\end{table}

\begin{figure*}[ht!]
	\begin{center}
		\includegraphics[width=.8\hsize]{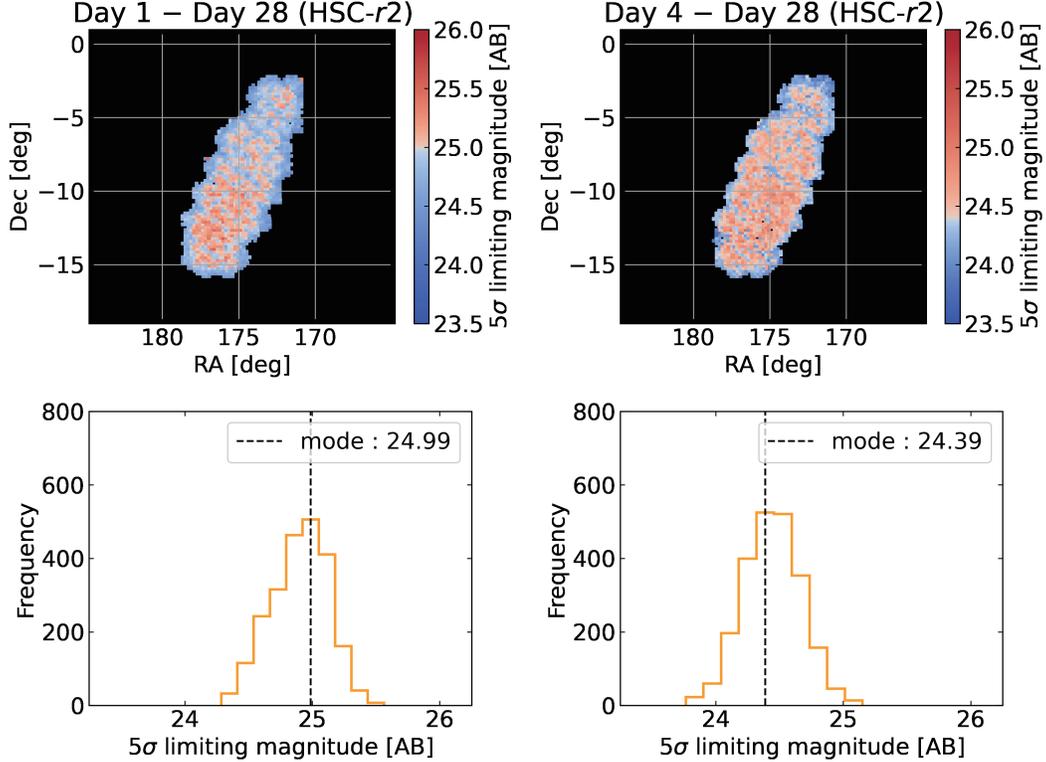} 
	\end{center}
	\caption{Map (top) and histogram (bottom) of the $5\sigma$ limiting magnitudes in the difference images within the first (Day~1 $-$ Day~28; left) and second epochs (Day~4 $-$ Day~28; right) in the $r2$-band. The color indicates larger/smaller (in red/blue) values than the mode value of each epoch.}
	\label{fig:limmag-HSC-R2}
\end{figure*}

We performed image subtraction for the science images obtained in the first (one day from GW detection; Day~1) and second (Day~4) epochs using the images taken in the third (Day~28) epoch as reference images.
The subtraction package is included in hscPipe v4.0.5 and is based on an algorithm in which the FWHM of the reference images is fitted to that in the science images via convolution using kernels to make their PSFs equivalent, as proposed in \citet{1998ApJ...503..325A} and \citet{1999astro.ph..3111A}.
Table \ref{tab:limmag_subim} lists the mode, maximum, and minimum values of the $5\sigma$ limiting magnitudes evaluated from the difference images.
The positional dependence of the $5\sigma$ limiting magnitudes in the $r2$-band is shown in Figure \ref{fig:limmag-HSC-R2}.

\begin{table}[ht!]
	\caption{Mode, maximum, and minimum values of the $5\sigma$ limiting magnitudes in the difference images.}
	\centering
	\begin{tabular}{lcccc}
        \hline \hline
        Difference Image & Filter & \multicolumn{3}{c}{Limiting magnitude (AB)} \\
                         &        & mode & max & min                            \\
        \hline
        Day 1$-$28 & $r2$   & 24.99 & 25.56 & 24.29                       \\
                         &  $z$   & 23.28 & 23.89 & 22.54                       \\
        Day 4$-$28 & $r2$   & 24.39 & 25.15 & 23.76                       \\
                         &  $z$   & 22.97 & 23.57 & 22.23                       \\
        \hline
	\end{tabular}
	\label{tab:limmag_subim}
\end{table}

\subsection{Source detection and screening} \label{ssec:screening}
\begin{figure}[ht!]
	\begin{center}
		\includegraphics[width=1.0\hsize]{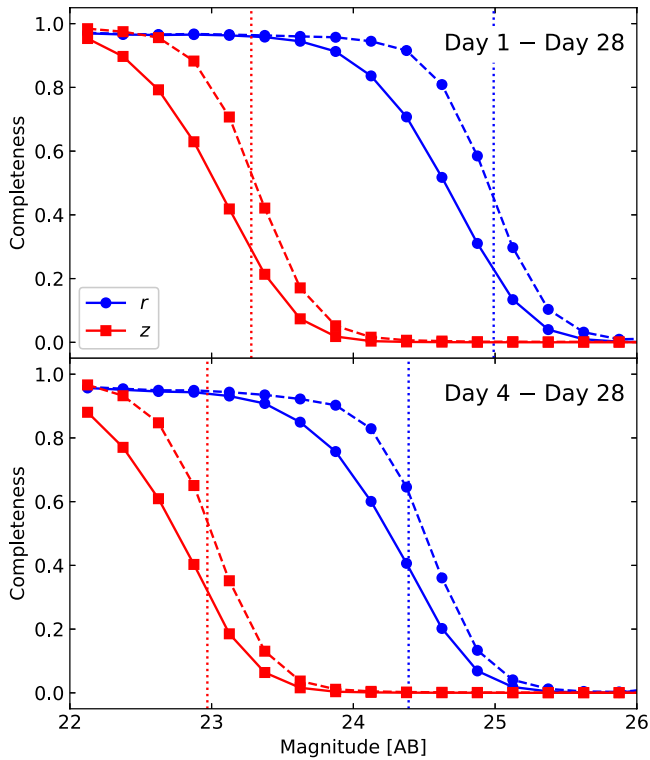} 
	\end{center}
	\caption{Completeness of transient detection in the difference images (top: Day~1$-$28; bottom: Day~4$-$28). The dashed curves are the detection fractions of artificial point sources as a function of magnitude. The solid curves are the detection fractions of the remaining artificial point sources after applying Criteria (i)$-$(iv). The vertical dotted lines indicate the mode values of the $5\sigma$ limiting magnitudes.}
	\label{fig:completeness}
\end{figure}

After image subtraction, we applied the following criteria to exclude bogus detections (for example, caused by bad pixels and failure of image subtraction) and select point sources from the difference images as in \citet{2018PASJ...70...28T} and \citet{2021PASJ...73..350O}: (i) A signal-to-noise ratio of PSF flux $(S/N)_{\rm PSF}>5$; (ii) $(b/a)/(b/a)_{\rm PSF}>0.65$, where $a$ and $b$ are the lengths of the major and minor axes, respectively, of the shape of a source; (iii) $0.7<{\rm FWHM}/({\rm FWHM})_{\rm PSF}<1.3$; (iv) PSF-subtracted residual with $<3\sigma$ standard deviation in the difference image.
Furthermore, we imposed the following criterion to exclude moving objects such as minor planets: (v) Detection at least twice in the difference images.
As a result, we obtained 5213 variable point sources.
To evaluate the completeness of transient detection with image subtraction and our detection criteria, we randomly injected artificial point sources with various magnitudes into the observed images and detected them in the difference images using the same detection criteria.
Figure \ref{fig:completeness} shows the completeness of transient detection in the difference images within the first (top: Day~1$-$28) and second epochs (bottom: Day~4$-$28).
The vertical dotted lines indicate the mode values of the $5\sigma$ limiting magnitudes, and these magnitudes were approximately comparable to the magnitude for a completeness of 23\% (Day~1$-$28, $r$), 29\% (Day~1$-$28, $z$), 39\% (Day~4$-$28, $r$), and 32\% (Day~4$-$28, $z$).

\begin{figure*}[ht!]
	\begin{center}
		\includegraphics[width=1.0\hsize]{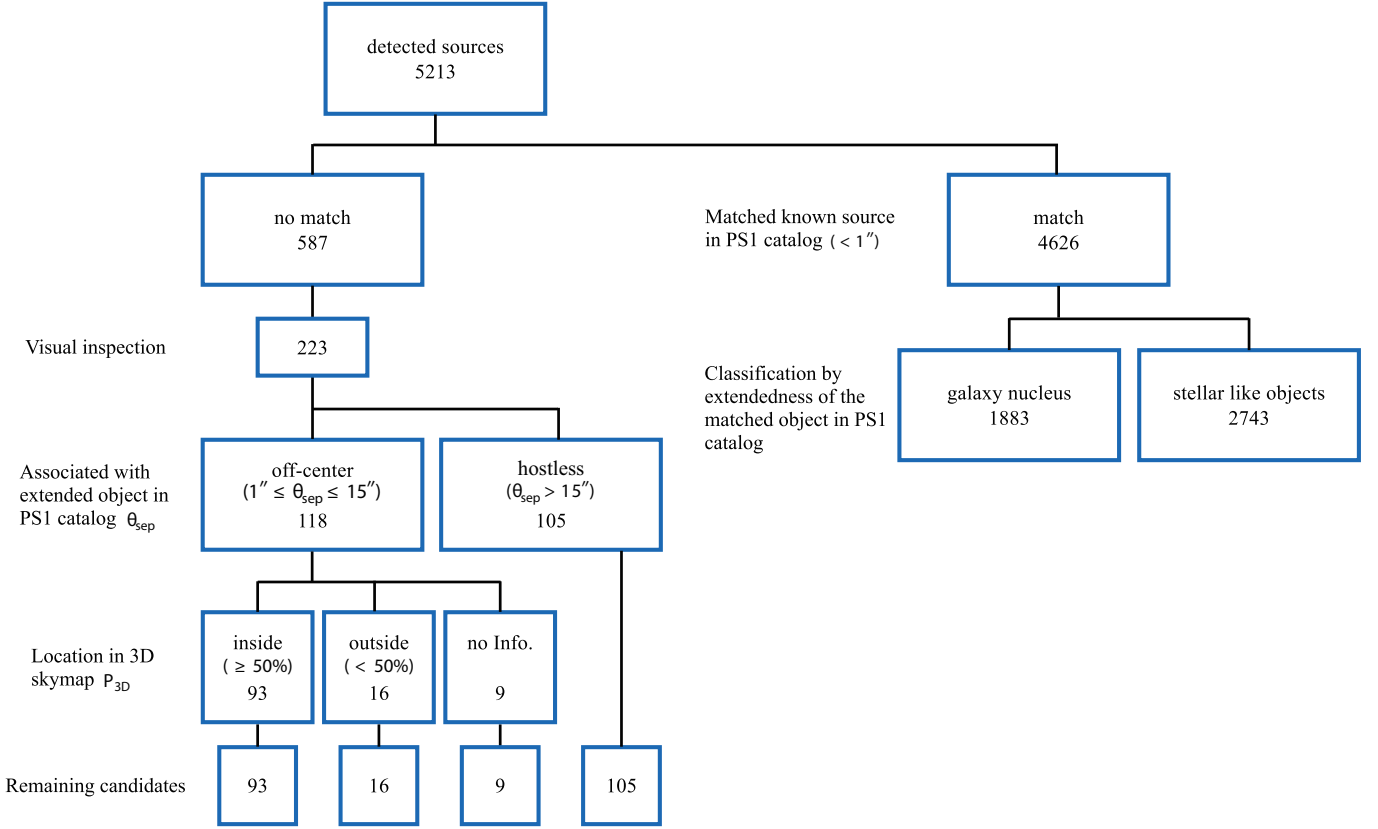}
	\end{center}
	\caption{Flowchart of the candidate screening and classification process for the detected sources after image subtraction. The number in each box represents the number of remaining sources after each process.}
	\label{fig:flowchart}
\end{figure*}

Figure \ref{fig:flowchart} shows a flowchart of the candidate screening and classification process for the detected sources.
We first checked whether the sources were associated with
known objects (for example, variable stars and AGNs) by
matching these sources with the Pan-STARRS1 (PS1) catalog
\citep{2020ApJS..251....7F} within 1$''$. Furthermore, to classify the PS1
objects, we used a flag, \texttt{objInfoFlag}, that indicates if an object is extended or not. As a result, we found that 2743
and 1883 sources were associated with stellar-like objects and extended
objects, respectively. The former are likely to have originated from
stellar variabilities, and thus we excluded them as candidates of an
EM counterpart of GW200224\_222234. The latter could be
variabilities of AGNs, an EM signal from a
BBH coalescence, or something else. The rate of BBH coalescence may be enhanced in galactic nuclei
\citep[for example,][]{2019ApJ...884L..50M,2020ApJ...899...26T,2020ApJ...898...25T}. 
The light curves of AGNs are stochastic \citep[for example,][]{2004ApJ...601..692V},
and thus a clear difference between AGN variability and a catastrophic
phenomenon such as BBH coalescence is the long-term variability. However, it was not possible to distinguish them from only
our three-nights’ observation over one month. Therefore, we also excluded the 1883 sources associated
with extended objects in this study. Further examination is left to
future studies.
After this selection, 587 sources remained in the sample.

As the next step, we performed a visual inspection to exclude
bogus detections and obtained 223 plausible candidates. Then, we classified these candidates by whether there are extended objects in the PS1 catalog within an angular separation, $\thetasep$.
Here, we classified them into two groups: ``off-center'' candidates
($1''\leq\thetasep\leq15''$) and ``hostless'' candidates
($\thetasep>15''$). We conservatively adopted the threshold ($\thetasep=15''$)
corresponding to a separation of $\sim60$~kpc at the distance of
GW200224\_222234 ($z\sim0.3$). This separation is larger than
the typical size of galaxies \citep[for example,][]{2003MNRAS.343..978S,2003MNRAS.341...54K}.
Then, we investigated off-center candidates with large separation individually. 
As a result, we obtained 118 off-center
candidates and 105 hostless candidates.

Furthermore, we calculated the probability $P_{\rm 3D}$ for 118
off-center candidates. $P_{\rm 3D}$ represents how likely the extended objects were to be located inside the 3D skymap of GW200224\_222234.
The definition and calculation method are described in \citet{2018PASJ...70...28T}.
We used the $r$- and/or $i$-band Kron-magnitudes (\texttt{rMeanKronMag}, \texttt{iMeanKronMag} in the PS1 catalog) of the extended objects to derive the absolute magnitudes of probable host galaxy in the observer frame.
We classified them with a threshold of $P_{\rm 3D}=0.5$ and obtained
93 likely candidates within the
$3\sigma$ error of the arrival distance of GW200224\_222234
($P_{\rm 3D}\geq0.5$) and 16 candidates that were likely outside ($P_{\rm 3D}<0.5$).
The remaining nine candidates had no information of the
Kron-magnitudes in the $r$- and $i$-bands in the PS1 catalog
(\texttt{rMeanKronMag}~$=$~\texttt{iMeanKronMag}~$=-999$) and were classified
as ``No Info.'' 
In the following, we investigate the nature of 93 candidates inside the 3D skymap, 16 candidates
outside the 3D skymap, nine candidates without host galaxy information, and
105 hostless candidates.

\section{Light curve fitting} \label{sec:LCFitting}
\subsection{Method} \label{ssec:method}

We performed light curve fitting using transient templates and examples to 
exclude common transients such as supernovae (SNe) from the 223 candidates.
We adopted a template set including the transient templates of Type Ia SNe
\citep{2007ApJ...663.1187H} and core-collapse SNe \citep[CCSNe; Type
Ibc, IIP, IIL, and IIn,][]{2002PASP..114..803N}, and examples of rapid transients \citep[RTs,][]{2014ApJ...794...23D} as in \citet{2018PASJ...70..103T}.
RTs were included in the template set despite their
low rate \citep{2014ApJ...794...23D} because their nature is still under debate
and they could have a similar timescale to a possible EM
counterpart of BBH coalescence \citep{2019ApJ...884L..50M}. The
template set did
not include other types of transients, such as superluminous SNe
\citep{2021MNRAS.504.2535I} and SNe amplified by a gravitational lens
\citep{2010MNRAS.405.2579O}, because there is no
indication that they are associated with BBH coalescence, and their rates are sufficiently low
that no detection was expected in our observation
\citep[for example,][]{2014Sci...344..396Q,2019ApJS..241...16M}.
Note that the probability that the nature of detected candidates is that of such transients cannot be excluded.

We derived the light curves of 223 candidates via forced PSF photometry of the difference 
images at their location. The difference fluxes of the candidates were 
measured using the difference images with the local background subtracted. 
Note that the difference fluxes could not be the genuine fluxes of candidates
because the time interval between the science and reference images was shorter 
than the typical timescales of SNe. Thus, forced PSF photometry was also performed 
for the stacked images without the local background subtracted. The measured fluxes 
included the fluxes of their host galaxies as well as the genuine fluxes of candidates
and thus were adopted as the upper limits of the genuine fluxes of candidates.
Then, we performed template fitting not only to the difference fluxes but also to 
the upper limits.

We considered a variation in the explosion date and
an intrinsic variation in templates and examples.
The light curve templates were derived with an explosion date $t_{\rm exp}$, redshift $z$, and variations as conducted in \citet{2018PASJ...70..103T}.
To account for the variations in supernova properties, we derived the peak absolute $B$-band magnitude $M_B$ using Equation (4) in \citet{2012ApJ...745...31B} by considering the stretch $s$, color $c$, and intrinsic variation $I$ for the template of a Type Ia SN.
For CCSNe and RTs, we parameterized $M_B$ and the color excess of their host galaxy $E_{B-V}$ characterizing the extinction of the host galaxy.
We assumed that the extinction curve of the host galaxy was the same as that in our Galaxy \citep{1992ApJ...395..130P}.

To evaluate the difference between the observed light curves and the
template, we defined the following $\xi$ value:
\begin{eqnarray}
	\xi := && \sum_{i}^{N^{\rm obs}_{\rm d}}\frac{\left( f^{\rm obs}_{{\rm d}, i}-f^{\rm temp}_{{\rm d}, i}(p) \right)^2}{{\sigma^{\rm obs}_{{\rm d}, i}}^2} \nonumber \\
	&& + \sum_{j}^{N^{\rm obs}_{\rm s}}\frac{\left( f^{\rm obs}_{{\rm s}, j}-f^{\rm temp}_{{\rm s}, j}(p) \right)^2}{{\sigma^{\rm obs}_{{\rm s}, j}}^2},
	\label{eq:chi2}
\end{eqnarray}
where the subscripts ${\rm d}$ and ${\rm s}$ indicate quantities
derived from the difference images (d) and stacked
images (s), respectively,
$N^{\rm obs}$ is the number of data points of the observation,
$f^{\rm obs}$ and $\sigma^{\rm obs}$ are the observed flux and its error, respectively, and $f^{\rm temp}(p)$ is the template flux calculated using a template parameter set $p$.
When $f^{\rm obs}_{{\rm s}, j}-f^{\rm temp}_{{\rm s}, j} >0$, we set $\sigma^{\rm obs}_{\rm s}$ to infinity because the flux in the stacked image is an upper limit of the genuine flux of the candidate.

The $5\sigma$ limiting magnitudes in the difference images correspond to 0.37\,$\mu{\rm Jy}$ (Day~1$-$28) and 0.64\,$\mu{\rm Jy}$ (Day~4$-$28) in the $r2$-band flux, and 1.77\,$\mu{\rm Jy}$ (Day~1$-$28) and 2.36\,$\mu{\rm Jy}$ (Day~4$-$28) in the $z$-band flux.
Most of these candidates were detected in the $r2$-band but not in the $z$-band.
If the difference flux was lower than the flux corresponding to the $5\sigma$ limiting magnitudes, we required the template flux to be fainter than the flux corresponding to the $5\sigma$ limiting magnitudes.

We applied the Metropolis--Hastings (MH) algorithm, which is a type of Markov-chain Monte Carlo (MCMC) method, for the template fitting.
This method can derive best-fit template parameter sets without being trapped by the local minima of $\xi$ by probabilistically moving to a location where $\xi$ is smaller in a parameter space.
We applied the top-hat function as a prior probability distribution with the same parameter ranges as adopted in \citet{2018PASJ...70..103T}.
The ranges were obtained from \citet{2012ApJ...745...31B} for the Type Ia SN, \citet{2012ApJ...757...70D} for CCSNe, and an assumption of $\pm1$ mag variation for RTs, as shown in Table \ref{tab:prior_range}.
We derived a Markov chain using 50000~samples with each redshift from 0.00 to 0.50 every 0.05 using each template model (Type Ia SN, 6 CCSNe, and 9 RTs).
Each sample of the Markov chain was proposed from a Gaussian distribution with the standard deviation shown in Table \ref{tab:prior_range}, using the current value as the median.
The initial values of each parameter were set to the central value of each range. We adopted a large number of burn-in steps to ensure a negligible dependence on the initial values. 
To narrow down possible templates, we selected samples with a loose
threshold of $\xi<25$ from all MCMC samples.

\startlongtable
\begin{deluxetable}{lcc}
\tablecaption{Template parameters applied in the MCMC method. \label{tab:prior_range}}
\tabletypesize{\scriptsize}
\tablehead{
\colhead{Parameter} &
\colhead{Range$^\dag$} & 
\colhead{SD of sampling$^\ddag$}
}
\startdata
\multicolumn{3}{c}{Common} \\
\hline
Explosion Time $t_{\rm exp}$ (day)    & $[-\infty,~\infty]$ & $0.1$  \\
\hline
\multicolumn{3}{c}{Type Ia SN} \\
\hline
Color $c$ & $[-0.2,~0.8]$ & $0.01$ \\
Stretch $s$ & $[0.6,~1.2]$ & $0.01$ \\
Intrinsic Variation $I$ & $[-0.3,~0.3]$ & $0.01$ \\
\hline
\multicolumn{3}{c}{Type IIL normal} \\
\hline
Peak Magnitude $M_{B}$ & $[-16.09,~-18.05]$ & $0.01$ \\
Color Excess $E_{B-V}$ & $[0,~1]$ & $0.01$ \\
\hline
\multicolumn{3}{c}{Type IIL bright} \\
\hline
Peak Magnitude $M_{B}$ & $[-17.92,~-19.96]$ & $0.01$ \\
Color Excess $E_{B-V}$ & $[0,~1]$ & $0.01$ \\
\hline
\multicolumn{3}{c}{Type IIP} \\
\hline
Peak Magnitude $M_{B}$ & $[-14.43,~-18.91]$ & $0.01$ \\
Color Excess $E_{B-V}$ & $[0,~1]$ & $0.01$ \\
\hline
\multicolumn{3}{c}{Type IIn} \\
\hline
Peak Magnitude $M_{B}$ & $[-16.98,~-20.66]$ & $0.01$ \\
Color Excess $E_{B-V}$ & $[0,~1]$ & $0.01$ \\
\hline
\multicolumn{3}{c}{Type Ibc normal} \\
\hline
Peak Magnitude $M_{B}$ & $[-16.09,~-18.05]$ & $0.01$ \\
Color Excess $E_{B-V}$ & $[0,~1]$ & $0.01$ \\
\hline
\multicolumn{3}{c}{Type Ibc bright} \\
\hline
Peak Magnitude $M_{B}$ & $[-18.46,~-20.30]$ & $0.01$ \\
Color Excess $E_{B-V}$ & $[0,~1]$ & $0.01$ \\
\hline
\multicolumn{3}{c}{Rapid Transient (PS1-10ah)} \\
\hline
Peak Magnitude $M_{B}$ & $[-16.63,~-18.63]$ & $0.01$ \\
Color Excess $E_{B-V}$ & $[0,~1]$ & $0.01$ \\
\hline
\multicolumn{3}{c}{Rapid Transient (PS1-10bjp)} \\
\hline
Peak Magnitude $M_{B}$ & $[-17.20,~-19.20]$ & $0.01$ \\
Color Excess $E_{B-V}$ & $[0,~1]$ & $0.01$ \\
\hline
\multicolumn{3}{c}{Rapid Transient (PS1-11bbq)} \\
\hline
Peak Magnitude $M_{B}$ & $[-18.48,~-20.48]$ & $0.01$ \\
Color Excess $E_{B-V}$ & $[0,~1]$ & $0.01$ \\
\hline
\multicolumn{3}{c}{Rapid Transient (PS1-11qr)} \\
\hline
Peak Magnitude $M_{B}$ & $[-18.03,~-20.03]$ & $0.01$ \\
Color Excess $E_{B-V}$ & $[0,~1]$ & $0.01$ \\
\hline
\multicolumn{3}{c}{Rapid Transient (PS1-12bb)} \\
\hline
Peak Magnitude $M_{B}$ & $[-15.29,~-17.29]$ & $0.01$ \\
Color Excess $E_{B-V}$ & $[0,~1]$ & $0.01$ \\
\hline
\multicolumn{3}{c}{Rapid Transient (PS1-12bv)} \\
\hline
Peak Magnitude $M_{B}$ & $[-18.44,~-20.44]$ & $0.01$ \\
Color Excess $E_{B-V}$ & $[0,~1]$ & $0.01$ \\
\hline
\multicolumn{3}{c}{Rapid Transient (PS1-12brf)} \\
\hline
Peak Magnitude $M_{B}$ & $[-17.39,~-19.39]$ & $0.01$ \\
Color Excess $E_{B-V}$ & $[0,~1]$ & $0.01$ \\
\hline
\multicolumn{3}{c}{Rapid Transient (PS1-13ess)} \\
\hline
Peak Magnitude $M_{B}$ & $[-17.49,~-19.49]$ & $0.01$ \\
Color Excess $E_{B-V}$ & $[0,~1]$ & $0.01$ \\
\hline
\multicolumn{3}{c}{Rapid Transient (PS1-13duy)} \\
\hline
Peak Magnitude $M_{B}$ & $[-17.77,~-19.77]$ & $0.01$ \\
Color Excess $E_{B-V}$ & $[0,~1]$ & $0.01$
\enddata
\tablecomments{[$\dag$] Range of top hat distribution for prior distribution.
[$\ddag$] Standard deviation of the proposed distribution for sampling.}
\end{deluxetable}

Additionally, for the candidates associated with the PS1 extended object (that is, off-center candidates), we confirmed the consistency between the redshifts of transient templates and the distance information of the host galaxies by applying the following criteria:
\begin{description}
    \item[Criterion 1-1] We eliminated templates whose redshifts were outside the redshift error range ($2\sigma$) of the PS1 extended object.
    \item[Criterion 1-2] If the PS1 objects exhibited a large angular separation, $\thetasep > 10''$, we did not apply Criterion 1-1 because the associated PS1 objects were potentially not true host galaxies.
     \item[Criterion 2] When $\xi>25$ for $\thetasep < 10''$ with Criterion~1-1, we allowed the templates at a redshift larger than that of the PS1 extended objects because a true faint host galaxy may exist behind the PS1 extended object even if $\thetasep < 10''$.
\end{description}

\begin{figure}[ht!]
    \begin{center}
	    \includegraphics[height=.3\textheight]{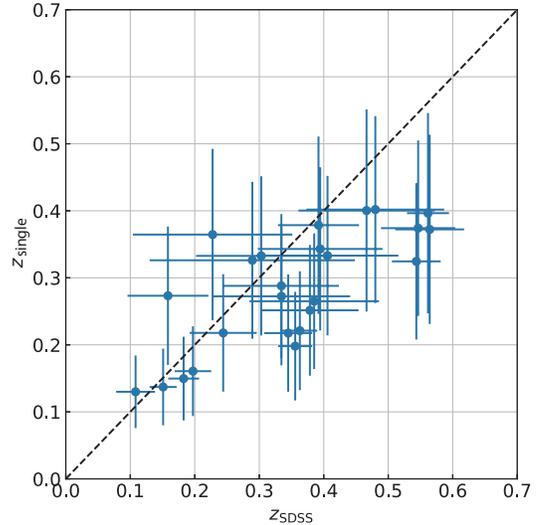}
	\end{center}
	\caption{Comparison between the SDSS photometric redshifts and the redshift estimated with Eqs.~(\ref{eq:mean_redshift}) and (\ref{eq:std_redshift}).}
	\label{fig:compare_redshift}
\end{figure}

Here, we used the photometric redshifts (photo-$z$) from the Sloan Digital Sky Survey \citep[SDSS;][]{2000AJ....120.1579Y} catalog as the redshift of the PS1 extended object if available.
If not, we estimated a single-band redshift $z_{\rm single}$ and its standard deviation $\sigma_{z}$ using the following formulae:
\begin{eqnarray}
	z_{\rm single} &:=& \frac{\int^{\infty}_{0} \phi\, A\,z(D)~dD}{\int^{\infty}_{0}\phi~A~dD}, \label{eq:mean_redshift} \\
	\sigma_{z}^2 &:=& \frac{\int^{\infty}_{0} \phi\, A\,(z(D)-z_{\rm single})^2~dD}{\int^{\infty}_{0}\phi~A~dD},
	\label{eq:std_redshift}
\end{eqnarray}
where $\phi = \phi(\lambda,~M)$ is the luminosity function of galaxies at a rest wavelength $\lambda$ derived from the luminosity functions in the $UBVRI$-band provided in \citet{2005A&A...439..863I} and {\it Planck} cosmology \citep{2014A&A...571A..16P}, $A = A(D)$ is a surface area observed at a distance $D$, $M = M(D;~m_{j})$ is an absolute magnitude derived with the $j$th-band ($r$ or $i$) apparent magnitude $m_{j}$ at $D$ in the observer frame, and $\lambda = \lambda(D;~\lambda_{j})$ is the rest wavelength redshifted from an observed wavelength $\lambda_{j}$ at $D$.
We also estimated $z_{\rm single}$ and $\sigma_{z}$ for
candidates with the SDSS photometric redshift
(Table~\ref{tab:consistent_cand}). Their comparison illustrates their approximate consistency (Figure~\ref{fig:compare_redshift}).
Here, we corrected the Galactic extinction
\citep{2011ApJ...737..103S}\footnote{\url{http://irsa.ipac.caltech.edu/applications/DUST/}}
when we derived absolute magnitudes from apparent
magnitudes. 

\subsection{Results of fitting analysis} \label{ssec:fitting results}
\begin{figure}[ht!]
    \begin{center}
	    \includegraphics[width=0.9\hsize]{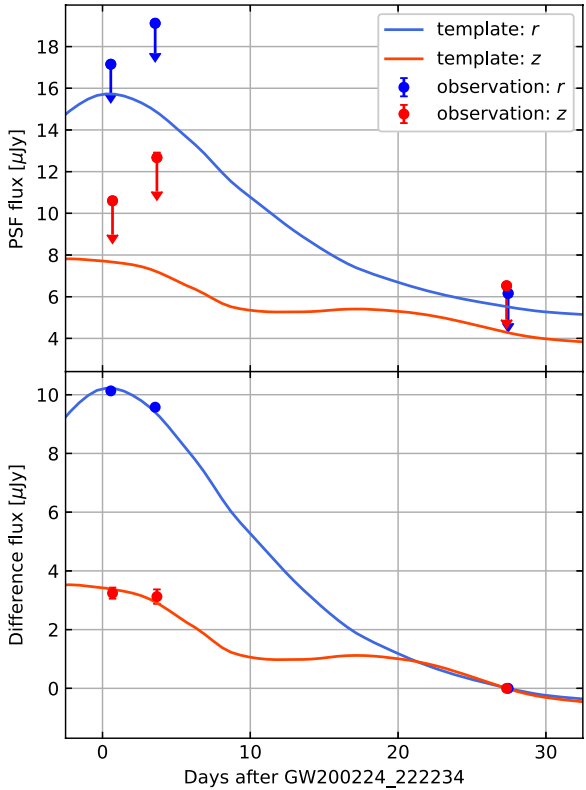}
	\end{center}
	\caption{Example of the fitting results. The light curve of JGEM20ewa (points with error bars), and the best-fit light curve of the Type Ia SN template (solid curves); $z=0.25$, $t_{\rm exp}=0.38$~d, $c=-0.15$, $s=0.60$, and $I=-0.02$. The vertical axes of the top and bottom panels indicate the upper limits measured in the stacked images and the difference fluxes measured in the difference images, respectively. }
	\label{fig:lightcurve_JGEM20ewa}
\end{figure}

Among the 223 candidates for which we performed light curve fitting,
200 candidates were consistent with
the templates of SNe, three candidates were consistent only with the
template of RTs, and the remaining 20 candidates were not consistent with any templates or examples.
Table \ref{tab:consistent_cand} shows the coordinates, $\thetasep$,
$P_{\rm 3D}$, and probable transient templates of the
200 candidates consistent with the templates.
A fitting result of JGEM20ewa with $\xi=0.6$ is shown as an example of the candidates consistent with the transient template (Figure \ref{fig:lightcurve_JGEM20ewa}).
We concluded that the most probable origin of this candidate is a Type Ia SN for a redshift of $z=0.25$.
The best-fit light curves are shown as solid lines in Figure \ref{fig:lightcurve_JGEM20ewa}.
The top and bottom panels show the PSF flux measured in the stacked and difference images, respectively.

\begin{deluxetable*}{lccccccrrl}
\tablecaption{Information of the candidates that are consistent with the transient templates or examples. \label{tab:consistent_cand}}
\tablewidth{0pt}
\tabletypesize{\scriptsize}
\tablehead{
\colhead{Name} &
\multicolumn{2}{c}{Coordinate (J2000.0)} &
\colhead{$\thetasep^{\dag}$} &
\colhead{$z_{\rm single}$} &
\colhead{$\sigma_{z}$} &
\colhead{$z_{\rm SDSS}$} &
\colhead{$\sigma_{\rm SDSS}$} &
\colhead{$P_{\rm 3D}^{\dag}$} &
\colhead{Probable transient template sets} \\
\colhead{} &
\colhead{R.A. (HH:MM:SS.ss)} &
\colhead{Decl. (DD:MM:SS.s)} &
\colhead{($''$)} &
\colhead{} &
\colhead{} &
\colhead{} &
\colhead{} &
\colhead{} &
\colhead{}
}
\startdata
JGEM20abe & 11:43:12.36 & $-$15:39:25.6 & 8.4 & 0.37 & 0.13 & -- & -- & 77 & CCSNe (Ibc and IIL) \\
JGEM20abf & 11:43:11.52 & $-$15:28:16.3 & -- & -- & -- & -- & -- & -- & CCSNe (Ibc, IIn and IIL) \\
JGEM20acf & 11:41:48.24 & $-$14:55:29.3 & -- & -- & -- & -- & -- & -- & CCSNe (Ibc, IIn and IIL) \\
JGEM20adp & 11:50:40.82 & $-$15:14:02.8 & -- & -- & -- & -- & -- & -- & Type Ia, CCSNe (Ibc, IIn and IIL) \\
JGEM20adq & 11:50:42.55 & $-$15:13:57.4 & -- & -- & -- & -- & -- & -- & CCSNe (Ibc, IIn and IIL) \\
JGEM20ads & 11:50:52.06 & $-$15:10:58.4 & 2.4 & 0.15 & 0.06 & -- & -- & 97 & Type Ia, CCSNe (Ibc, IIn, IIP and IIL) \\
JGEM20aej & 11:49:46.03 & $-$15:01:31.8 & -- & -- & -- & -- & -- & -- & Type Ia, CCSNe (Ibc, IIn, IIP and IIL), RTs (13duy) \\
JGEM20afe & 11:48:22.75 & $-$15:44:46.3 & 12.9 & 0.33 & 0.12 & -- & -- & 93 & CCSNe (Ibc and IIL) \\
JGEM20cvb & 11:40:58.56 & $-$11:26:57.5 & 2.0 & -- & -- & -- & -- & -- & Type Ia \\
JGEM20ewa & 11:37:13.22 & $-$8:21:04.7 & 3.6 & 0.15 & 0.06 & -- & -- & 82 & Type Ia \\
$\vdots$ & $\vdots$ & $\vdots$ & $\vdots$ & $\vdots$ & $\vdots$ & $\vdots$ & $\vdots$ & $\vdots$ & $\vdots$ \\
\enddata
\tablecomments{This table is published in its entirety in the machine-readable format. A portion is shown here for guidance regarding its form and content.
[$\dag$] We indicate ``--'' if a candidate is not associated with PS1 extended objects.}
\end{deluxetable*}

Among the 200 candidates consistent with the templates of SNe, 105 candidates were associated with the PS1 extended objects (that is, off-center candidates), and the remaining 95 candidates had no PS1 extended objects within $<15''$.

\begin{figure}[ht!]
	\begin{center}
		\includegraphics[width=1.0\hsize]{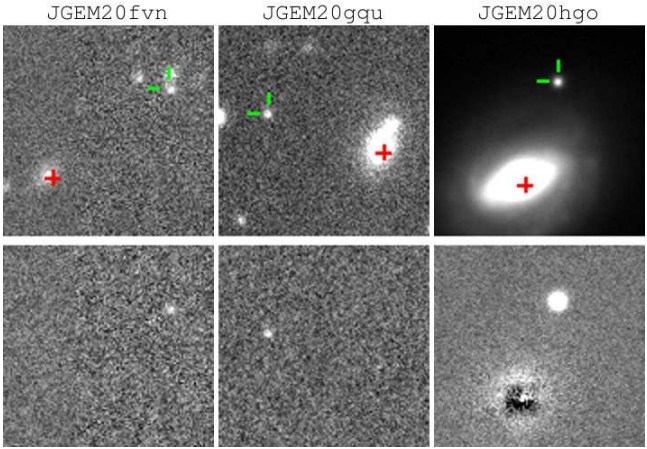}
	\end{center}
	\caption{Observed (top) and difference (bottom) $r2$-band images of JGEM20fvn, JGEM20gqu, and JGEM20hgo on February 25, 2020. Each image displays the $20''\times20''$ region. Although the PS1 extended objects (plus-shaped markers) are associated with the candidates (reverse L-shaped markers), these are unlikely to be the true host galaxies of the candidates.}
	\label{fig:noHosts_thumbnail}
\end{figure}

Criterion~1-1 was applied to 102 out of the 105 off-center candidates, and these candidates were excluded from the final candidates.
Two of the remaining candidates (JGEM20fvn and JGEM20gqu) were associated with the PS1 extended objects of $z=0.56\pm0.05$ and $0.34\pm0.04$, respectively.
The positions of the candidates and PS1 objects are shown
in Figure \ref{fig:noHosts_thumbnail}; left (JGEM20fvn) and middle
(JGEM20gqu). Because JGEM20fvn and JGEM20gqu exhibited $\thetasep=13''.8$ and $11''.0$ respectively, Criterion~1-2 was applied to them.
We found that the templates of Type Ibc CCSNe with $z=0.2$ and $0.15$ reproduced the light curves of JGEM20fvn and JGEM20gqu, respectively.
These results indicated that the PS1 objects associated with JGEM20fvn
and JGEM20gqu were unlikely to be true host galaxies and were likely to be Type Ibc CCSNe with $z=0.2$ and $0.15$. Thus, we ruled out these two objects from the final candidates.
Criterion~2 was applied to the remaining one candidate (JGEM20hgo). With Criterion~2,
we found that a template of Type Ibc CCSN with $z=0.1$ can reproduce the light curve of JGEM20hgo.
The redshift of the template was larger than the redshift of the PS1 object associated with JGEM20hgo ($z=0.03\pm0.01$).
Therefore, we also ruled out JGEM20hgo from the final candidates because it could be a Type Ibc CCSN behind the PS1 object.
The stacked and difference images of JGEM20hgo and the associated PS1 object are shown in right panels of Figure \ref{fig:noHosts_thumbnail}.
The parameters $\thetasep$ and $P_{\rm 3D}$ of candidates consistent with the transient templates are shown in Table \ref{tab:consistent_cand}.

\begin{figure}[ht!]
    \begin{center}
	    \includegraphics[width=0.9\hsize]{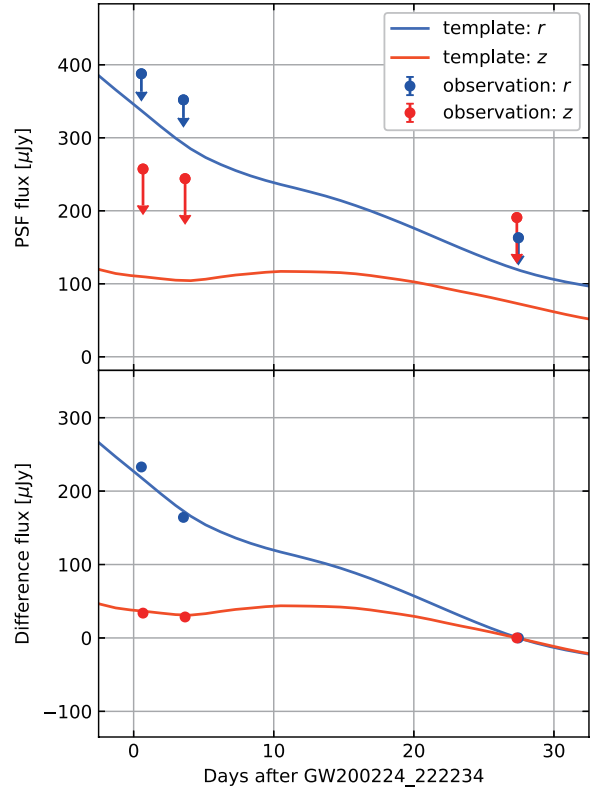}
	\end{center}
	\caption{Light curve of JGEM20cvb (points with error bars), and the best-fit light curve of the Type Ia SN template (solid curves); $z=0.05$, $t_{\rm exp}=11.99$~d, ${\rm color}=-0.20$, $s=1.07$, and $I=0.05$.}
	\label{fig:lightcurve_JGEM20cvb}
\end{figure}

Among the 20 candidates inconsistent with the templates, one candidate (JGEM20cvb) exhibited a large $\xi$ value of 2175.
However, it apparently matched the template light curve, as shown in Figure \ref{fig:lightcurve_JGEM20cvb}.
This large $\xi$ could be attributed to the underestimation of photometric errors or the uncertainties on the template fluxes.
For instance, $\xi$ was improved to $\sim2.4$ by assuming a template difference flux error of 0.1 mag.
Thus, we concluded that the origin of JGEM20cvb is a Type Ia SN and excluded it from the final candidates.

We also performed template fitting without the
single-band photometric redshifts for the candidates associated with
extended objects without SDSS photometric redshifts. We found that the
redshifts of the best-fit templates were consistent for most of the
candidates and confirmed that the consistent templates did not
change for candidates other than the candidates discussed above, that is, JGEM20fvn, JGEM20gqu, and JGEM20hgo.

\begin{figure*}[ht!]
    \begin{center}
	    \includegraphics[height=.92\textheight]{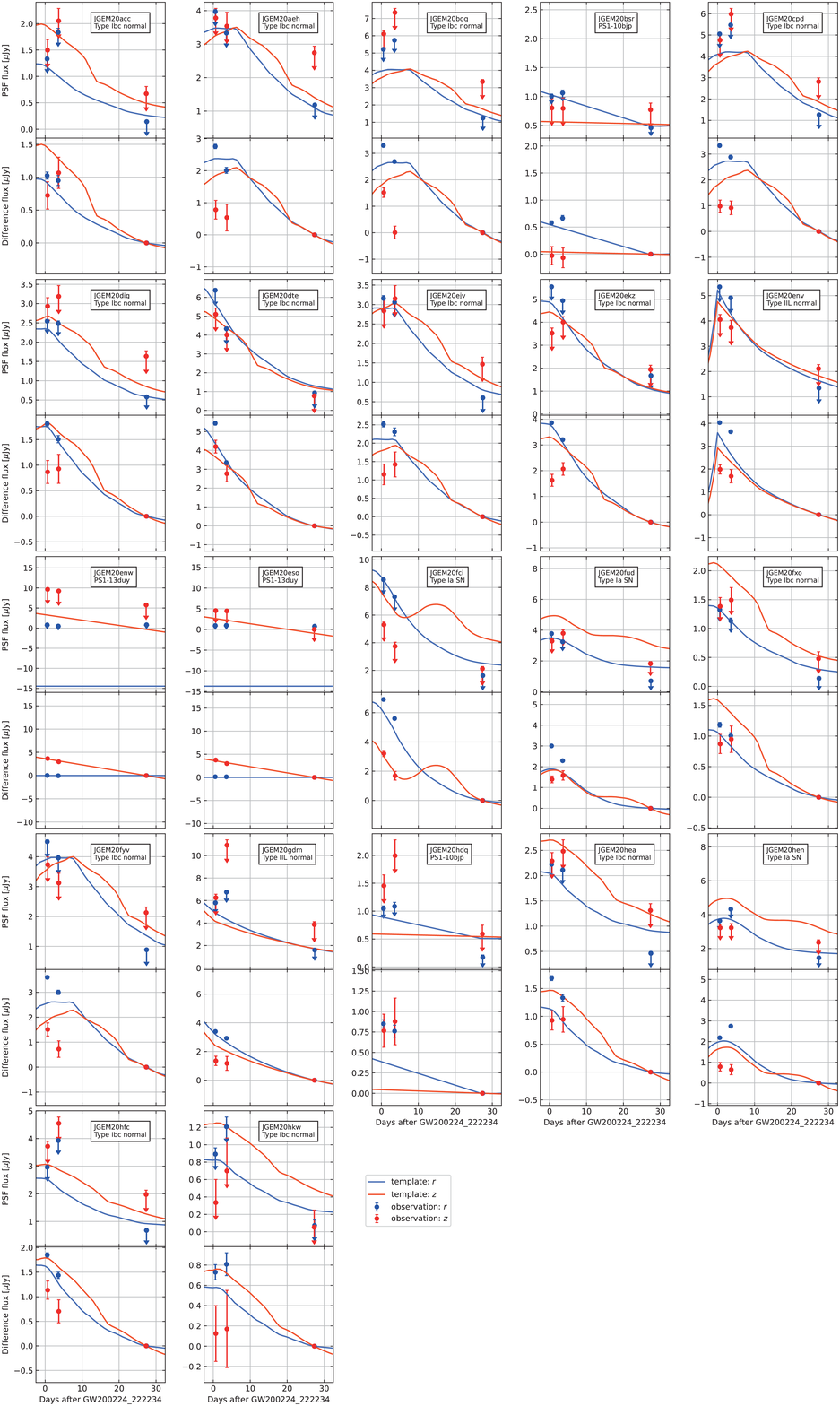}
	\end{center}
	\caption{Light curves of 19 candidates inconsistent with the templates of SNe and three candidates consistent only with RTs. The best-fit (but bad-fit for the former 19 candidates) templates or examples are also shown for comparison. The vertical axis of each panel indicates the upper limits measured in the stacked images (top) and the difference fluxes measured in the difference images (bottom).}
	\label{fig:inconsistent_lightcurve}
\end{figure*}

As a result, 22 objects, that is, three objects consistent only with RTs and 19
objects inconsistent with all templates and examples, remained as the final candidates of the optical counterpart of GW200224\_222234.
Their coordinates, $\thetasep$, $P_{\rm 3D}$, and fitting results are shown in Table \ref{tab:inconsistent_cand} and Figure \ref{fig:inconsistent_lightcurve}.

\begin{deluxetable*}{lccrrl}
\tablecaption{Information on candidates inconsistent with the templates of SNe and their associated PS1 extended objects. \label{tab:inconsistent_cand}}
\tablewidth{0pt}
\tablehead{
\colhead{Name} &
\multicolumn{2}{c}{Coordinate (J2000.0)} &
\colhead{$\thetasep^{\dag}$} &
\colhead{$P_{\rm 3D}^{\dag}$} &
\colhead{Best fitted transient templates} \\
\colhead{} &
\colhead{R.A. (HH:MM:SS.ss)} &
\colhead{Decl. (DD:MM:SS.s)} &
\colhead{($''$)} &
\colhead{} &
\colhead{}
}
\startdata
JGEM20acc & 11:41:52.70 & $-$15:08:22.9 & -- & -- & CCSN Type Ibc ($\xi\sim33$) \\
JGEM20aeh & 11:50:20.04 & $-$15:12:54.0 & -- & -- & CCSN Type Ibc ($\xi\sim55$) \\
JGEM20boq & 11:39:19.68 & $-$12:44:10.0 & 8.9 & 75 & CCSN Type Ibc ($\xi\sim334$) \\
JGEM20bsr & 11:37:26.83 & $-$12:05:34.4 & -- & -- & RT 10bjp ($\xi\sim16$) \\
JGEM20cpd & 11:35:05.23 & $-$11:06:33.8 & 14.9 & 94 & CCSN Type Ibc ($\xi\sim281$) \\
JGEM20dig & 11:41:48.72 & $-$11:45:02.9 & 11.4 & 78 & CCSN Type Ibc ($\xi\sim26$) \\
JGEM20dte & 11:27:56.54 & $-$9:05:45.6 & -- & -- & CCSN Type Ibc ($\xi\sim159$) \\
JGEM20ejv & 11:47:24.98 & $-$9:40:32.5 & 6.3 & 75 & CCSN Type Ibc ($\xi\sim62$) \\
JGEM20ekz & 11:45:32.81 & $-$9:42:04.7 & -- & -- & CCSN Type Ibc ($\xi\sim84$) \\
JGEM20env & 11:33:13.03 & $-$8:10:35.0 & 14.3 & -- & CCSN Type IIL ($\xi\sim331$) \\
JGEM20enw & 11:32:50.98 & $-$7:56:35.2 & 14.9 & 74 & RT 13duy ($\xi\sim1$) \\
JGEM20eso & 11:28:42.14 & $-$7:48:41.4 & -- & -- & RT 13duy ($\xi\sim10$) \\
JGEM20fci & 11:44:18.89 & $-$8:08:34.4 & -- & -- & Type Ia SN ($\xi\sim1137$) \\
JGEM20fud & 11:31:23.86 & $-$5:55:16.0 & 7.0 & 89 & Type Ia SN ($\xi\sim1562$) \\
JGEM20fxo & 11:40:19.39 & $-$6:36:34.9 & -- & -- & CCSN Type Ibc ($\xi\sim77$) \\
JGEM20fyv & 11:40:00.36 & $-$6:01:27.5 & 5.2 & 99 & CCSN Type Ibc ($\xi\sim422$) \\
JGEM20gdm & 11:45:52.56 & $-$6:13:49.4 & 1.6 & 91 & CCSN Type IIL ($\xi\sim27$) \\
JGEM20hdq & 11:32:35.04 & $-$3:56:15.7 & 3.0 & 60 & RT 10bjp ($\xi\sim187$) \\
JGEM20hea & 11:32:20.45 & $-$3:40:50.9 & 3.7 & 60 & CCSN Type Ibc ($\xi\sim394$) \\
JGEM20hen & 11:32:21.91 & $-$3:09:21.6 & 7.3 & 98 & Type Ia SN ($\xi\sim348$) \\
JGEM20hfc & 11:31:30.55 & $-$3:45:54.7 & 2.7 & 94 & CCSN Type Ibc ($\xi\sim87$) \\
JGEM20hkw & 11:35:25.49 & $-$4:11:36.6 & -- & -- & CCSN Type Ibc ($\xi\sim37$)
\enddata
\tablecomments{[$\dag$] We indicate ``--'' if a candidate is not associated with PS1 extended objects.}
\end{deluxetable*}

\section{Spectroscopic observations with the GTC/OSIRIS} \label{sec:GTC observation}
Optical spectroscopic observations were performed using the OSIRIS instrument mounted on the 10.4-\,m GTC telescope located at the Roque de los Muchachos Observatory (Canary Islands, Spain).
Five targets were selected for spectroscopic follow-ups to determine the spectroscopic redshift (spec-$z$) of the probable host galaxies of the candidates.
The selected targets were the PS1 extended objects associated with JGEM20fud, JGEM20fyv, JGEM20gdm, JGEM20hdq, and JGEM20hen, which were sufficiently bright in the $z$-band for short exposures.
Owing to observation time constraints, we targeted only these five objects.

The observations were performed on February 10, 2021. The instrumental setup was the spectroscopic long-slit mode, with a R1000R grism and slit width of 1 arcsec.
Each target observation was divided into three exposures of 400\,s (1200\,s in total for JGEM20fud, JGEM20fyv, and JGEM20hen) and 600\,s (1800\,s in total for JGEM20gdm and JGEM20hdq).
In addition to the targets, the spectrophotometric standard star G191-B2B was also observed using the same instrumental setup and observation conditions.
Standard calibration images for bias, flat field, and the calibration lamp (HgAr+Xe+Ne) were also taken during the same night.
The data were reduced using standard procedures for bias subtraction, flat-field correction, flux calibration, and atmospheric extinction correction.

The flux calibrated spectra are shown in Figures \ref{fig:osiris1} and \ref{fig:osiris2}.
The spec-$z$ was derived based on the identification of spectral lines.
In addition to the redshifted spectral features from the galaxies, we identified a possible intervening system in the line of sight of JGEM20fud and JGEM20hen, which is consistent with an Mg absorption line located at $z=0.33$.
We show the coordinate, the $z$-band magnitude, estimated
redshift, and spec-$z$ of the target galaxies in Table
\ref{tab:result_GTC}. If an SDSS photometric redshift is not available, we
show $z_{\rm single}$ and $\sigma_{z}$ calculated with Equations
(\ref{eq:mean_redshift}) and (\ref{eq:std_redshift}). However, note that
these have large uncertainties because they are calculated from only a
single band magnitude.

\begin{figure*}[ht!]
    \begin{center}
	    \includegraphics[width=.95\hsize]{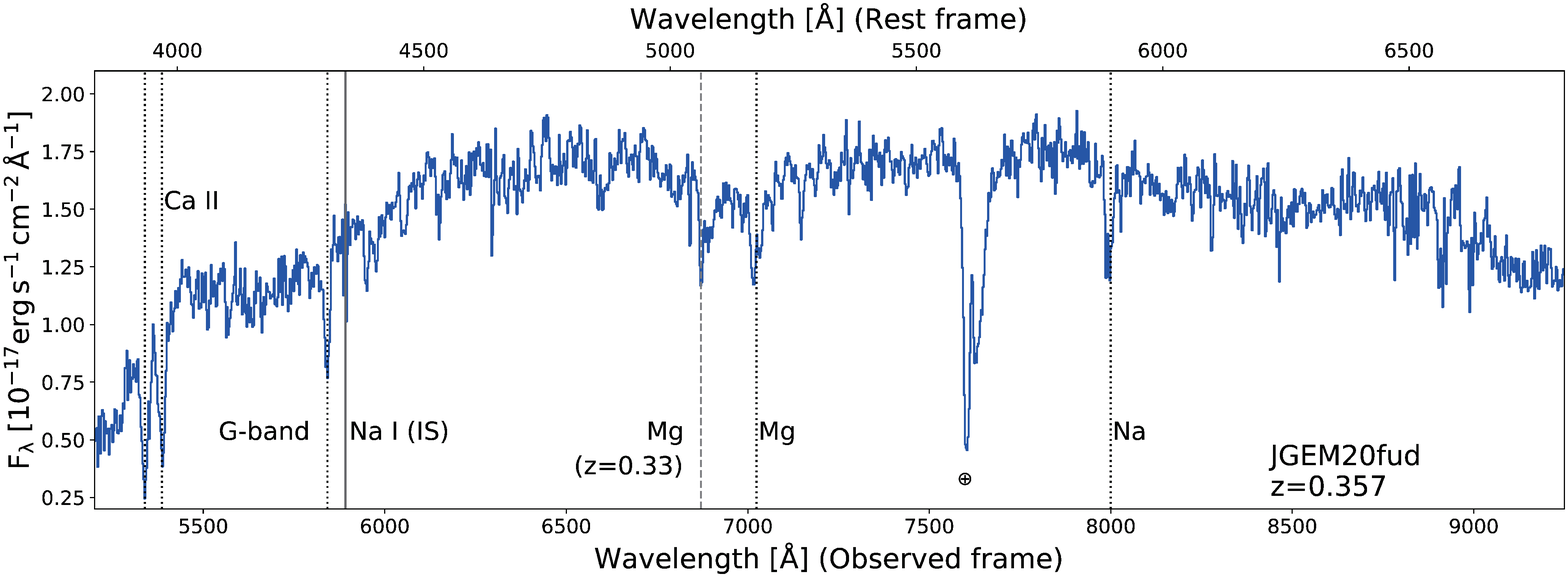}
	    \includegraphics[width=.95\hsize]{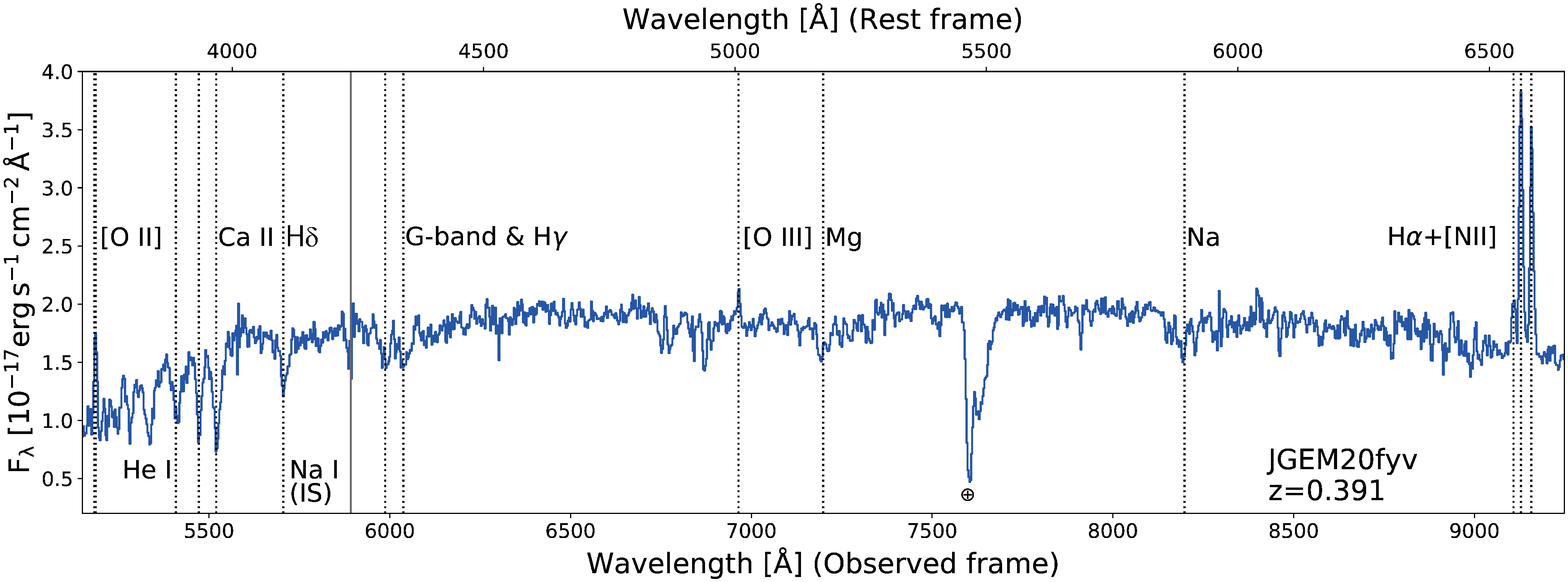}
	    \includegraphics[width=.95\hsize]{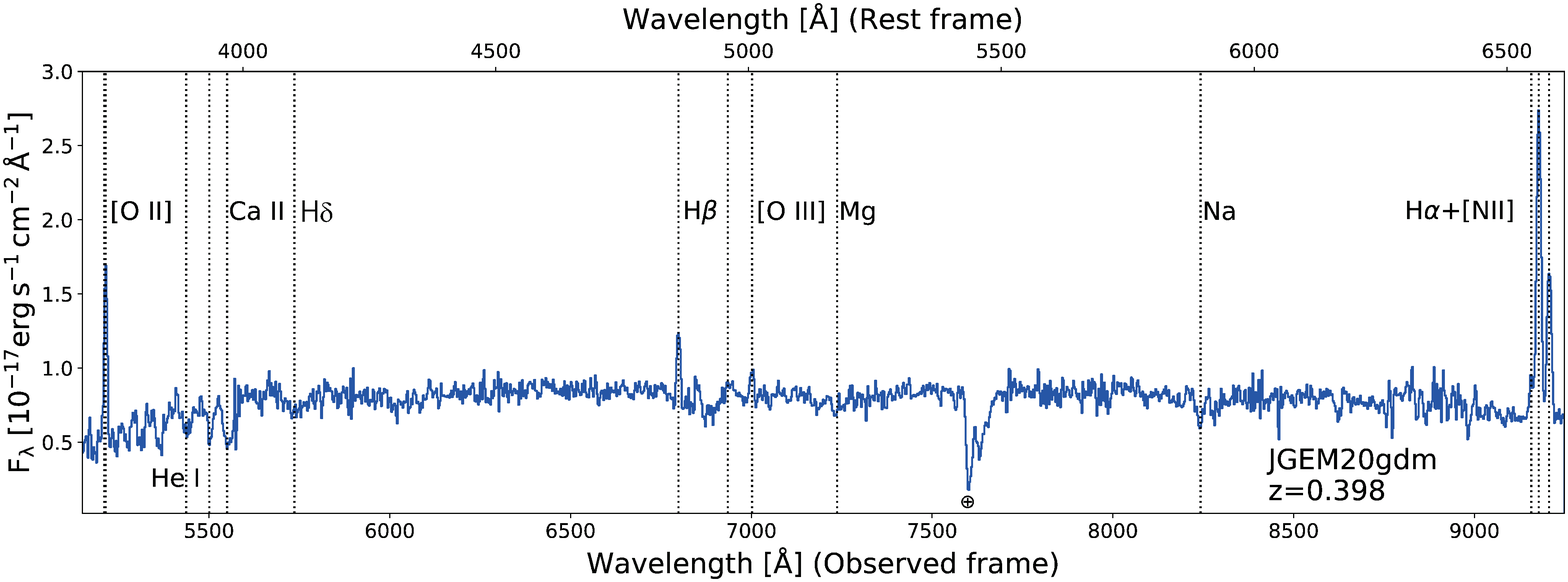}
	\end{center}
	\caption{Optical spectra of candidates JGEM20fud, JGEM20fyv, JGEM20gdm, JGEM20hdq, and JGEM20hen observed with GTC. The identified spectral features are marked by vertical dotted lines. The Na interstellar absorption feature is denoted by a solid vertical line. The emission lines identified as intervening systems are marked by dashed lines.}
	\label{fig:osiris1}
\end{figure*}

\begin{figure*}[ht!]
    \begin{center}
	    \includegraphics[width=.95\hsize]{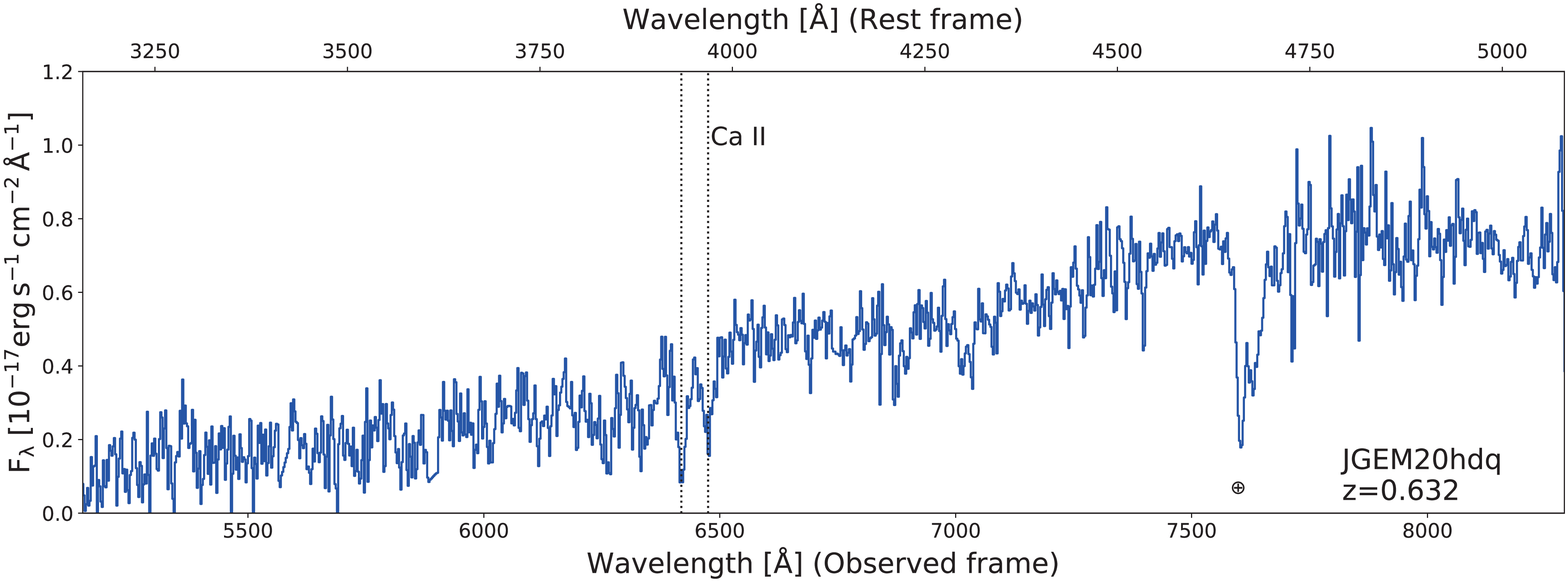}
	    \includegraphics[width=.95\hsize]{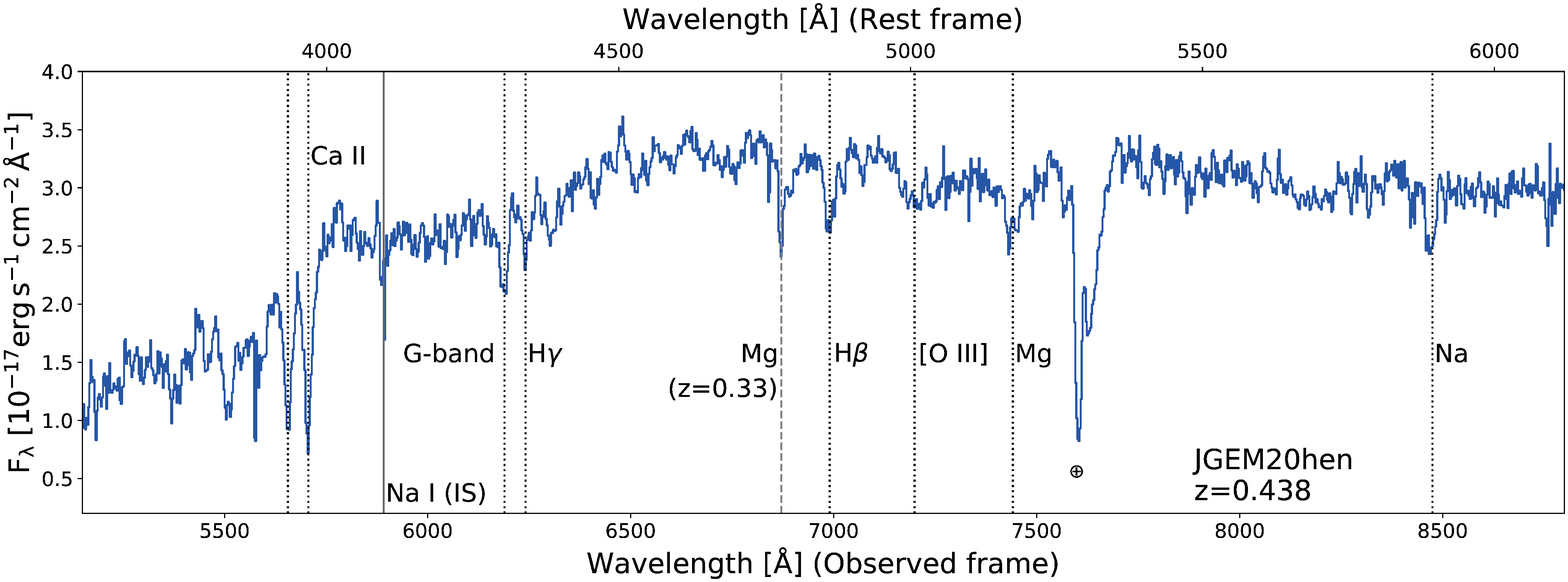}
	\end{center}
	\caption{Continuation of Figure \ref{fig:osiris1}.}
	\label{fig:osiris2}
\end{figure*}

\begin{figure*}[ht!]
	\begin{center}
	    \includegraphics[width=.8\hsize]{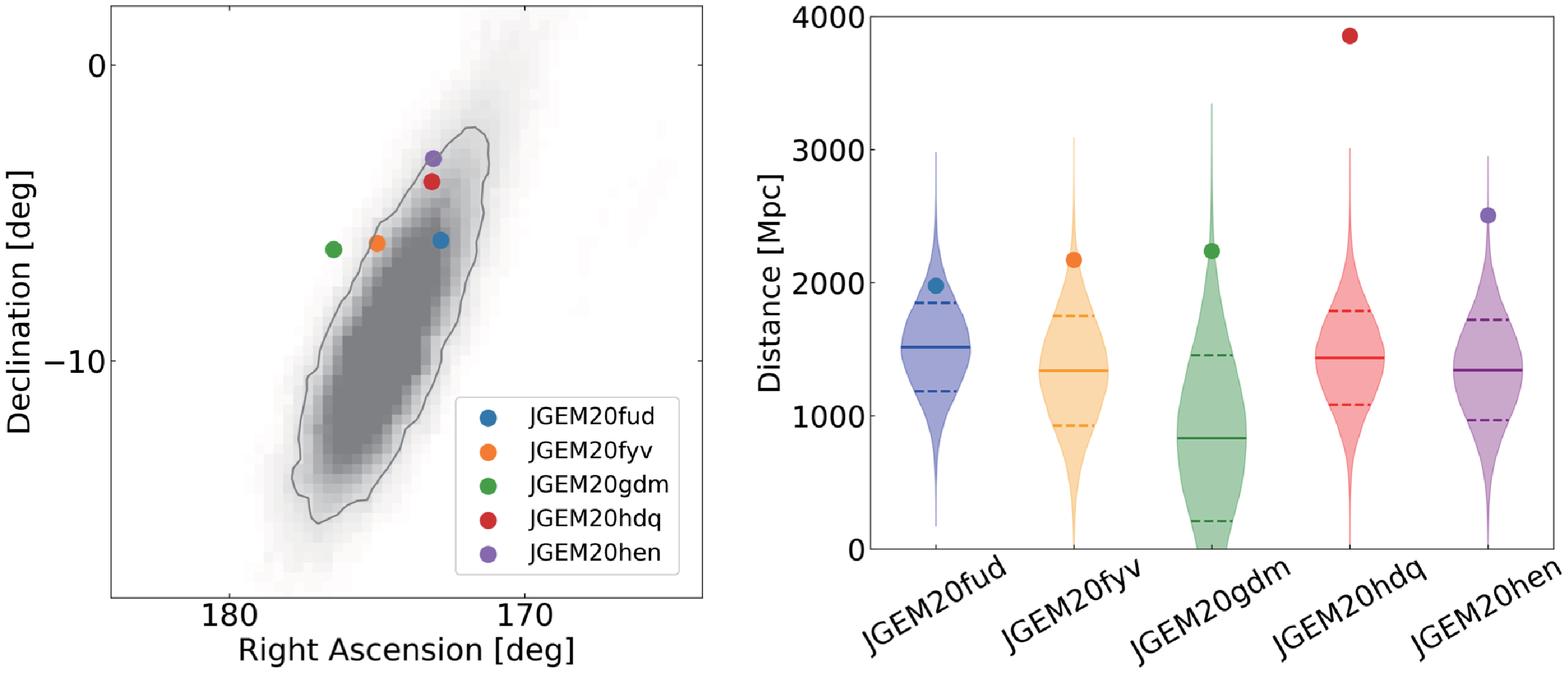}
	\end{center}
	\caption{(Left) 2D location of the five candidates observed with the GTC/OSIRIS and the localization skymap (IMRPhenomXPHM model) released in the GWTC-3 catalog. The dots are the locations of the candidates. The gray contour line indicates the 90\% credible region. (Right) Distance of each candidate and probability distribution at their location as a function of distance. The dots are the distances of the candidates. The horizontal solid and dashed lines indicate the mean and standard deviation of the probability distribution, respectively.}
	\label{fig:location of candidates}
\end{figure*}

Figure \ref{fig:location of candidates} shows the 2D location and distance of the five candidates.
We evaluated their associations with GW200224\_222234 by comparing the 3D location of their probable host galaxies with the 3D skymap of GW200224\_222234.
Here, we define the probability of the association as follows:
\begin{eqnarray}
	P_{\rm assoc}(x_{\rm gal}) := 1-\int_{f_{x}>f_{x_{\rm gal}}}f_{x}\,dV_{x},
	\label{eq:Passoc}
\end{eqnarray}
where $f_{x}$ is the probability distribution function (PDF) at the 3D coordinates $x$, and $x_{\rm gal}$ is the location of the host galaxy.
The integration in Equation (\ref{eq:Passoc}) represents the cumulative probability integrated over the region with $f_{x}>f_{x_{\rm gal}}$.
The PDF was obtained as $f_{x}=\rho({\rm R.A.},~{\rm Decl.})\,N(D;~\mu,~\sigma^2)$ using the 2D probability distribution $\rho=\rho({\rm R.A.},~{\rm Decl.})$ and normal distribution $N(D;~\mu,~\sigma^2)$ of the distance with the mean $\mu=\mu({\rm R.A.},~{\rm Decl.})$ and standard deviation $\sigma=\sigma({\rm R.A.},~{\rm Decl.})$ provided for each direction.
To perform the integration of Equation (\ref{eq:Passoc}) numerically, we discretized $N(D)$ into 5000~bins from 0 to 50000\,Mpc.
The discretization of $\rho$ was performed through pixelization using HEALPix.
$P_{\rm assoc}$ is small when the host galaxy is located outside the highly probable region.
We set the threshold to $P_{\rm assoc}=0.01$.
Table \ref{tab:result_GTC} also shows $P_{\rm assoc}$ obtained for the five targets.
Although JGEM20gdm, JGEM20hdq, and JGEM20hen were located outside the highly probable region ($P_{\rm assoc}<0.01$) and are unlikely to be related to GW200224\_222234,
JGEM20fud and JGEM20fyv were close to the highly probable region ($P_{\rm assoc}=0.76$ and $0.07$, respectively) and thus are possible counterparts of GW200224\_222234.

\begin{deluxetable*}{lcccccc}
\tablecaption{Information of the target galaxies. \label{tab:result_GTC}}
\tablewidth{0pt}
\tablehead{
\colhead{Name} &
\multicolumn{2}{c}{Coordinate of the target galaxy (J2000.0)} &
\colhead{${m_{z}}^\dag$} &
\colhead{Estimated} & 
\colhead{spec-$z$} &
\colhead{$P_{\rm assoc}$} \\
\colhead{} &
\colhead{R.A. (HH:MM:SS.ss)} &
\colhead{Decl. (DD:MM:SS.s)} &
\colhead{(mag)} &
\colhead{Redshift} &
\colhead{}
}
\startdata
JGEM20fud & 11:31:24.32 & $-$05:55:17.8 & 18.7 & $0.22\pm 0.03^\ddag$ & 0.357 & $0.76$ \\
JGEM20fyv & 11:40:00.36 & $-$06:01:22.1 & 19.0 & $0.26\pm 0.07^*$     & 0.391 & $0.07$ \\
JGEM20gdm & 11:45:52.63 & $-$06:13:48.4 & 19.9 & $0.29\pm 0.09^*$     & 0.398 & $1\times10^{-3}$ \\
JGEM20hdq & 11:32:35.03 & $-$03:56:18.6 & 20.1 & $0.56\pm 0.03^\ddag$ & 0.632 & $1\times10^{-8}$ \\
JGEM20hen & 11:32:22.41 & $-$03:09:22.9 & 18.4 & $0.36\pm 0.03^\ddag$ & 0.438 & $4\times10^{-3}$
\enddata
\tablecomments{[$\dag$] $z$-band magnitude of the target galaxy from the PS1 catalog. [$\ddag$] Photometric redshift $z_{\rm SDSS}$ from the SDSS catalog. [$^*$] $z_{\rm single}$ and $\sigma_{z}$ calculated with Eqs. (\ref{eq:mean_redshift}) and (\ref{eq:std_redshift}).}
\end{deluxetable*}

\section{Discussion} \label{sec:discussion}
\subsection{Contamination from supernovae} \label{ssec:SNrate}
Next, we evaluated our detection criteria and screening process by comparing our results with the expected number of SN detections.
We estimated the expected number of SN detections by summing up mock-SN samples weighted with cosmological histories of SN rates using the observation depth, as in \citet{2014PASJ...66L...9N} and \citet{2021PASJ...73..350O}.
We assumed the SN rate of \citet{2014PASJ...66...49O} and \citet{2012ApJ...757...70D} for the Type Ia SN and CCSN, respectively.
For the Type Ia SN, the SN light curves were generated from the evolution of the SN spectrum provided by \citet{2007ApJ...663.1187H}.
For the CCSN, the light curves were generated from the templates provided by \citet{2002PASP..114..803N}\footnote{The light curve templates of Type Ia SNe and CCSNe are available at the website \url{https://c3.lbl.gov/nugent/nugent\_templates.html}.}.
We referred to the luminosity distributions of SNe in \citet{2012ApJ...745...31B} and \citet{2012ApJ...757...70D}.

\begin{figure}[ht!]
	\begin{center}
		\includegraphics[width=1.\hsize]{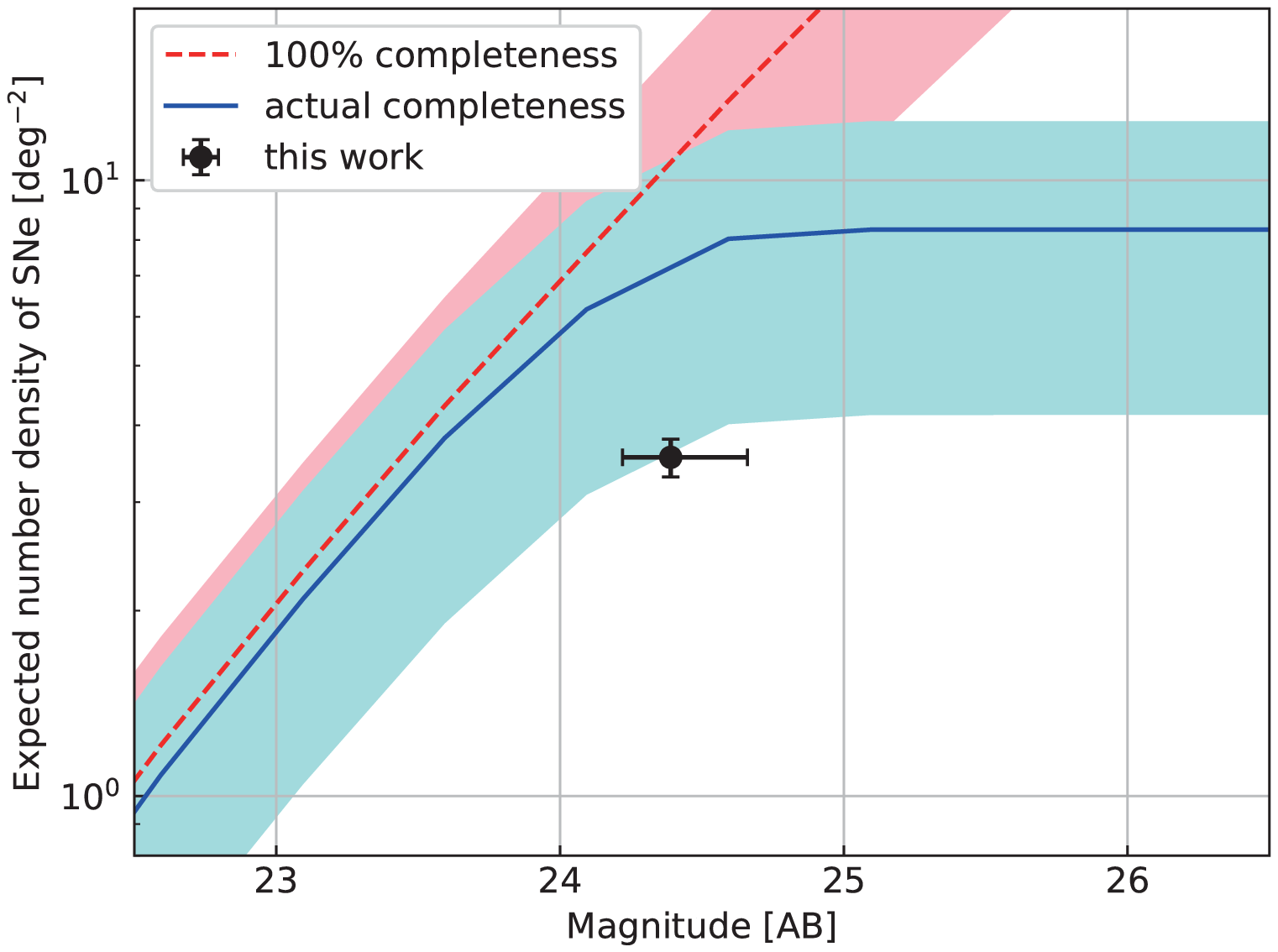} 
	\end{center}
	\caption{Expected number density of SN detection in the $r$-band as a function of limiting magnitude. The colored area is the $\pm 50\%$ error originating from the $1\sigma$ error of the core-collapse-SN rate density. The dashed curve was obtained with 100\% completeness, and the solid curve was obtained with the actual completeness of our survey, shown in Figure \ref{fig:completeness}, under the assumption of twice detection. The dot with error bars was obtained using the number density of the transients consistent with the templates of SNe. The vertical error bar is defined as $\sqrt{N_{\rm SN}}/S$ by assuming that the number follows a Poisson distribution with the expected value of $N_{\rm SN}$. The horizontal error bar is the $1\sigma$ error of the $5\sigma$ limiting magnitude in the difference images on Day~4 in the $r2$-band.}
	\label{fig:SNrate}
\end{figure}

Because our observation depth in the $z$-band was shallower than that in
The $r$-band, most of candidates were detected only in the $r$-band.
Thus, we sampled mock SNe whose brightness was decaying in the $r$-band,
assuming the reference images were taken 27~d after detection because the variation in the SN magnitude during 3~d between Days 1 and 4 ($<0.1$\,mag) was negligible compared to the standard deviation of the $5\sigma$ limiting magnitude of our observation.
The number density as a function of limiting magnitude is shown in Figure \ref{fig:SNrate} as a dashed curve.
The colored area represents the $\pm 50\%$ error from the $1\sigma$ error of the CCSN rate density.
In the actual observation, the detection completeness for fainter sources was lower (Figure \ref{fig:completeness}).
Therefore, we considered two detections based on the completeness of Day~1$-$28 and Day~4$-$28.
The solid curve in Figure \ref{fig:SNrate} is the number density derived by considering the completeness.
The dot with error bars was estimated with the number of transients consistent with SNe, $N_{\rm SN}=201$, and our observed area of $S=56.6$\,deg$^2$.
The vertical error bar was defined as $\sqrt{N_{\rm SN}}/S$ by assuming that the number followed a Poisson distribution with the expected value of $N_{\rm SN}$.
The horizontal error bar is the $1\sigma$ error of the $5\sigma$ limiting magnitude in the difference images on Day~4 in the $r2$-band.

Our result is consistent with the expected number density of SN
detection within a $1\sigma$ error. This verified that our detection
criteria and screening process can be used to identify SNe with high
completeness. Note that our result was located at the
$1\sigma$ lower end of the expected value. This might be because we ignored nuclear transients in this search.

\subsection{Implication in BBH merger models accompanied by EM emissions} \label{ssec:model implication}
We finally found 19 candidates of the optical counterpart of GW200224\_222234 by performing light curve fitting and spectroscopic observations.
Owing to the limitation of observations, we must exclude
sources located at the center of galaxies, and thus these final candidates cannot be considered variabilities of galactic nuclei, such as a kicked BBH merger in an accretion disk shown by \citet{2019ApJ...884L..50M}.
\citet{2016PTEP.2016e1E01Y} reported a BBH merger potentially accompanied by $\gamma$-ray emissions caused by a relativistic outflow.
They showed that optical afterglow peaks appear $\sim 4.5\times 10^4$\,s after GW detection (almost the same as the start of our first epoch observation) if the ambient matter density is 0.01\,cm$^{-3}$; however, the peak flux is expected to be lower than our observation limits (0.37\,$\mu$Jy in $r$-band) at a distance of 1710\,Mpc.
Therefore, such afterglow could not be detected in our search.

Among the 19 candidates, 16 candidates inconsistent with all templates and examples exhibited the increasing difference flux in one band and the decreasing difference flux in another band (for example, JGEM20ejv), or a more rapid decline in the $z$-band than in the $r$-band (for example, JGEM20boq).
If we perform further observations to obtain deeper images than the
current reference images after the candidates
sufficiently decay, we can measure
the genuine fluxes of these candidates instead of the difference
fluxes and hence identify their natures.

The final 19 candidates are potentially unrelated to GW200224\_222234.
If there are no counterparts of GW200224\_222234, the upper limits of optical luminosity are $\nu L_{\nu} < 5.2^{+2.4}_{-1.9}\times 10^{41}$ erg~s$^{-1}$ ($9.1^{+4.1}_{-3.3}\times 10^{41}$ erg~s$^{-1}$) and $\nu L_{\nu} < 1.8^{+0.8}_{-0.6}\times 10^{42}$ erg~s$^{-1}$ ($2.4^{+1.1}_{-0.9}\times 10^{42}$ erg~s$^{-1}$) on Day~1 (Day~4) in the $r2$-band and $z$-band, respectively.
Note that these upper limits are comparable with
the luminosity of a possible EM counterpart of the BBH
event GW190521 ($10^{42}$~erg~s$^{-1}$, \citealt{2019ApJ...884L..50M}).

\subsection{Comparison with sources detected by other observations} \label{ssec:Swift and DESGW}
For GW200224\_222234, the Neil Gehrels Swift Observatory performed near-UV/X-ray observations covering an area corresponding to 79.2\%/62.4\% of the GW probability region and reported their detection of eight X-ray sources and three near-UV sources \citep{Klingler2021, Oates2021}, whereas no significant signal was detected in very high-energy $\gamma$-ray \citep{2021ApJ...923..109A}.
We searched our sources to find out whether these sources were detected.
The source JGEM20aoz was located at ${\rm R.A.}=11^{\rm h}41^{\rm m}41^{\rm s}.81$, ${\rm Decl.}=-14^{\circ}07'54''.78$~(J2000.0), 4.8~arcsec off center of the error circle of X-ray Source~7.
JGEM20aoz was detected four times (two epochs and two bands) and exhibited a magnitude of 21.75\,mag (21.59\,mag) and 21.43\,mag (21.25\,mag) in the $r$- and $z$-bands, respectively, in the first (second) epoch difference images.
However, we classified it as a star-like object and ruled it out because the coordinates matched that of a point source found in the PS1 catalog.
In the SIMBAD archival database, a BL~Lac-type object (2FGL~J1141.7-1404) is located at a position consistent with JGEM20aoz.
Therefore, we conclude that JGEM20aoz is likely an AGN.

Our observation area also covered part of the area observed by the
GW program in the Dark Energy Survey (DESGW).
They conducted observations of GW200224\_222234 using the Dark Energy Camera (DECam) in the $i$-band on February 24, 25, 27, and March 5, 2020, with $10\sigma$ limiting magnitudes of 23.17, 23.19, 23.49, and 23.03, respectively, and reported eight transients \citep{2020GCN.27366....1M}.
Comparing the transients found by the DESGW with our candidates, JGEM20gfn matched the coordinates of AT2020ehw within $1''$.
JGEM20gfn was classified as an SN via light curve fitting.
We could not detect AT2020eho and AT2020eht because they were located at the CCD gap of our difference images.
Sources were detected at the location of AT2020ehp and AT2020ehq in our difference images, but they were rejected because of the large elongation, $(b/a)/(b/a)_{\rm PSF}=0.6$ and $0.5$, respectively.
Although no source was found at the position of AT2020ehy in both the stacked and difference images, AT2020ehv and AT2020ehr were observed in the stacked images but could not be detected in the difference images with our criterion $(S/N)_{\rm PSF}>5$.

\subsection{Comparison with a kilonova light curve} \label{ssec:kilonova}

Although GW200224\_222234 was classified as a BBH merger, we compared our
observation result with a kilonova model based on a radiative transfer simulation with an ejecta mass $M_{\rm ej}=0.05\,M_{\odot}$ and an electron fraction $Y_{\rm e}=0.30-0.40$ \citep{2020ApJ...901...29B}.
This kilonova model can explain the observed early multi-color light curve of GW170817/AT2017gfo.
The peak of bolometric luminosity approximately scales with $M_{\rm ej}$ to the power of 0.35 \citep[see][]{2016ARNPS..66...23F, 2019LRR....23....1M, 2016AdAst2016E...8T}.
$Y_{\rm e}$ influences the light curves through the opacity.
In the low-$Y_{\rm e}$ case ($Y_{\rm e}<0.25$), the kilonova ejecta becomes Lanthanide rich and the opacity becomes higher,
and thus the light curves in the early phase can be fainter by $2-3$ mag than that in the adopted model of $Y_{\rm e}=0.30-0.40$.

\begin{figure}
	\begin{center}
		\includegraphics[width=1.\hsize]{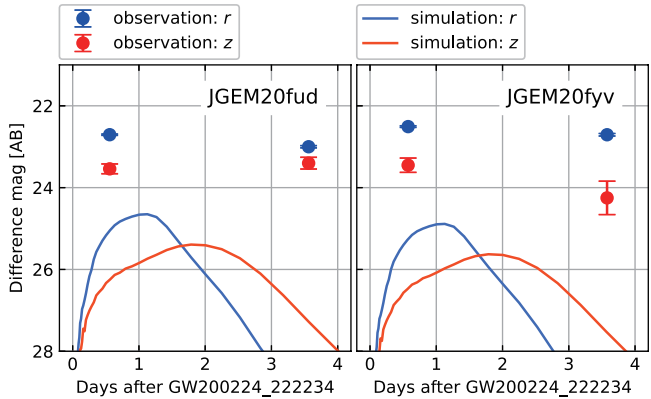} 
	\end{center}
	\caption{Comparison of light curves between the candidates whose host galaxies' spectroscopic redshifts are consistent with the 3D skymap of GW200224\_222234 and a kilonova model with an ejecta mass $M_{\rm ej}=0.05\,M_{\odot}$ and an electron fraction $Y_{\rm e}=0.30-0.40$ \citep{2020ApJ...901...29B}. We assume the kilonovae are located at the distance of each candidate. }
	\label{fig:CompGW170817Cand}
\end{figure}

Figure \ref{fig:CompGW170817Cand} shows a comparison between the kilonova light curve and the magnitudes of two candidates whose host galaxies' spectroscopic redshifts are consistent with the 3D skymap of GW200224\_222234 (JGEM20fud and JGEM20fyv).
All observed magnitudes were brighter and more slowly decaying than the kilonova at the distance of the host galaxies.
They did not exhibit rapid decline in short-wavelength components, being a major feature of the kilonova model; thus, they are unlikely to be kilonovae.

\subsection{Future prospects} \label{ssec:future prospects}

Here, we discuss the future prospects of follow-ups for
kilonova events with the Subaru/HSC in the era of
next-generation GW interferometers, such as an optimal upgrade of the
LIGO facilities, known as ``Voyager.''
If a BNS merger occurs at the distance of GW200224\_222234 ($\sim$1710~Mpc), can we detect it?

\begin{figure}
	\begin{center}
		\includegraphics[width=1.\hsize]{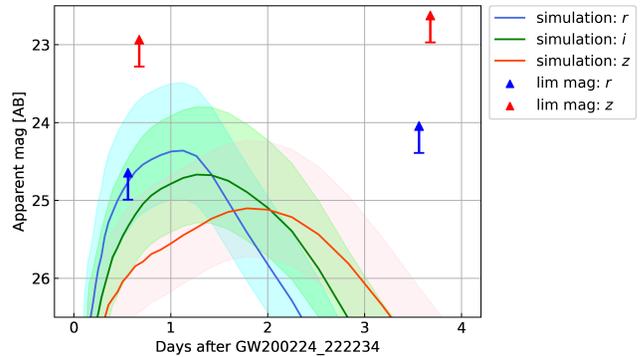} 
	\end{center}
	\caption{Comparison between a kilonova model located at the estimated distance of the GW200224\_222234 and the limiting magnitudes of our observation. The solid curves are the light curves of the kilonova model, the same as in Figure \ref{fig:CompGW170817Cand}. The colored area represents the uncertainty caused by the $1\sigma$ credible range of the distance.}
	\label{fig:CompGW170817}
\end{figure}

Figure \ref{fig:CompGW170817} shows the $r$-, $i$-, and $z$-band light curves under an assumption that the kilonova is located at the estimated distance of GW200224\_222234.
The colored area is the uncertainty caused by the 1$\sigma$ credible range of the distance.
The horizontal lines with the upward arrows are the 5$\sigma$ limiting magnitudes measured in the difference images in each epoch in each band ($r$ and $z$).
In the $r$-band, although the observation depth in the first epoch was sufficiently deep to detect the rising phase of the light curve, the model light curve decreased rapidly and was significantly fainter than the limiting magnitude in the second epoch.
The observation depths in the $z$-band did not reach the brightness in either epoch.
Thus, if a BNS merger occurs at the distance of GW200224\_222234, we might be able to detect the light curve in the early phase in the $r$-band using the power of a wide-field survey with an 8-\,m telescope.

Based on the follow-up of GW200224\_222234, the number of sources detected only on Day~1 and only in the $r2$-band (excluded using detection criterion (v)) reached 20137 \footnote{This number includes bogus detections because we did not perform visual inspection.}.
It is difficult to determine their origin using only
single epoch and single band data. 

Next, we consider what will improve the detection efficiency of future surveys. (1) If we
adopt a high-cadence (one day) and continuous (over three days) observation, the
kilonova at this distance will be detected on Days 1 and 2 and not detected
on Day 3 in the $r$-band. This will illustrate the rapidly
evolving nature of the kilonova. (2) If we adopt the $i$-band instead of the $z$-band, 
the $i$-band observation with Subaru/HSC will reach $\sim24$~mag with 1~min of exposure and
may detect the kilonova on Day 2. This will constrain the color of the
kilonova. These considerations illustrate
the power of a dedicated wide-field search with an 8m-class telescope, such as Subaru/HSC and Rubin/Legacy Survey of Space and Time (LSST).

\section{Summary \& Conclusion} \label{sec:summary and conclusion}
The BBH coalescence GW200224\_222234 was detected by the LIGO/Virgo collaboration with their three detectors on February 24, 2020.
We performed ToO observations using the Subaru/HSC in the $r2$- and $z$-bands during three epochs: February 24, 28, and March 23, 2020.
We selected the observation area from the high-probability region in the preliminary localization skymap covering 56.6\,deg$^2$.
The integrated probability reaches 91\% in the localization skymap released in the GWTC-3 catalog.

We searched for the optical counterpart using the image subtraction technique.
We adopted the images taken on the third epoch as the reference images and obtained the difference images of the first and second epochs.
After screening for the sources detected in the difference
images via matching with the PS1 catalog and visual inspection, we
found 223 candidates. We could not include sources located at the galactic center owing to the limitations of observation.
Subsequently, we classified these candidates using their angular separation from the nearby extended object and distance estimated from the photometric data.
Additionally, we investigated their nature using light curve fitting with the transient template set.
We adopted the templates of the Type Ia SN and CCSN and examples of RTs and then found 201 candidates consistent with the SN templates and likely not related to GW200224\_222234.

To measure the spectroscopic redshifts of the probable host galaxies of the final candidates, we also performed spectroscopic observations using the GTC/OSIRIS for the extended PS1 objects associated with the five candidates.
We found that two targets (JGEM20fud and JGEM20fyv) are likely to be located inside the highly probable region.
The other targets are outside the highly probable region and not related to the GW event.

As a result, we found 19 candidates as possible candidates of the optical counterpart of GW200224\_222234. The light curves of three candidates were consistent only with those of RTs, and the light curves of the other 16 candidates were inconsistent with all transients and examples.
These 19 candidates have a potential for being unrelated to GW200224\_222234; however, we could not establish their nature because of the lack of spectroscopic observations for the candidates.
If there is no counterpart of GW200224\_222234 in the 19 final candidates, the upper limits of optical luminosity are evaluated as $\nu L_{\nu} < 5.2^{+2.4}_{-1.9}\times 10^{41}$ erg~s$^{-1}$ ($9.1^{+4.1}_{-3.3}\times 10^{41}$ erg~s$^{-1}$) and $\nu L_{\nu} < 1.8^{+0.8}_{-0.6}\times 10^{42}$ erg~s$^{-1}$ ($2.4^{+1.1}_{-0.9}\times 10^{42}$ erg~s$^{-1}$) on Day~1 (Day~4) in the $r2$-band and $z$-band, respectively, from the $5\sigma$ limiting magnitudes of our observation.
These upper limits are comparable with the brightness of a
possible EM counterpart of the BBH event GW190521
($10^{42}$~erg~s$^{-1}$, \citealt{2019ApJ...884L..50M}).

We evaluated our detection criteria and screening process by comparing our result with the expected number of SN detections, which was estimated from a cosmic SN rate.
The number density of the candidates consistent with the SN templates was consistent with the expected number density within a $1\sigma$ error.
This indicated that our method could identify SNe with high completeness.

We also compared our sources with those found by the {\it Swift} and DESGW and found that some of our sources were associated with their sources. However, these identical sources are unlikely to be the optical counterpart of the GW.

Additionally, we discuss the implications of our result in several models of a BBH merger accompanied by EM emissions.
The kicked BBH merger in an accretion disk reported by
\citet{2019ApJ...884L..50M} cannot be a possible nature of the 16
candidates inconsistent with all transients because we excluded nucleus transients.
The relativistic outflow model reported in \citet{2016PTEP.2016e1E01Y} also cannot be a possible nature of the candidates because the expected flux is fainter than the $5\sigma$ limiting magnitude of our observations.
It is important to perform further observations deeper than the observations on Day~28 to identify their nature.

We also compared the light curve of the two candidates with the kilonova light curve by adopting a radiative transfer simulation that could explain the multi-color light curve of GW170817/AT2017gfo.
These candidates likely did not originate from the same kilonova as AT2017gfo because these were brighter and more slowly decaying than the kilonova.
We also discuss future prospects by comparing our observation depths with the kilonova light curve.
If we
perform a high-cadence (one day) and continuous (over three days)
observation, we can reveal the rapidly evolving nature of transients. Moreover, our observation can detect a kilonova or a possible EM
counterpart of the BBH event GW190521 \citep{2019ApJ...884L..50M} at the distance of GW200224\_222234. This demonstrates
the power of 8m-class telescopes, such as Subaru/HSC and Rubin/LSST.

\begin{acknowledgments}
These collaborative observations were possible thanks to the leadership of the National Astronomical Observatory of Japan (NAOJ) and the Instituto de Astrof\'{i}sica de Canarias (IAC).
This research was based in part on data collected at the Subaru Telescope, which is operated by the NAOJ, and in part on observations made using the GTC telescope in the Spanish Observatorio del Roque de los Muchachos of IAC under Director's Discretionary Time.
The spectroscopic data were obtained using the instrument OSIRIS, built by a Consortium led by the Instituto de Astrof\'isica de Canarias in collaboration with the Instituto de Astronom\'ia of the Universidad Aut\'onoma de M\'exico.
OSIRIS was funded by GRANTECAN and the National Plan of Astronomy and Astrophysics of the Spanish Government.
We are grateful to the staff of the Subaru Telescope for their help with the observations of this study.
We are honored and grateful for the opportunity to observe the Universe from Maunakea, which has cultural, historical, and natural significance in Hawaii.
This study was supported by MEXT KAKENHI (JP17H06363) and JSPS KAKENHI
 (JP19H00694, JP20H00158, JP20H00179, JP21H04997).
JBG, FP, JAAP, IPF, and TMD acknowledge financial support from the Spanish Ministry of Science and Innovation (MICINN) through the Spanish State Research Agency, under Severo Ochoa Program 2020-2023 (CEX2019-000920-S) and the projects PID2019-107988GB-C22 and PID2019-105552RB-C43.
We would like to thank Editage (www.editage.com) for English language editing.
\end{acknowledgments}

%

\facilities{Subaru (HSC), GTC(OSIRIS)}


\software{hscPipe \citep{2018PASJ...70S...5B},  
          astropy \citep{2013A&A...558A..33A,2018AJ....156..123A},  
          Source Extractor \citep{1996A&AS..117..393B}}






\bibliography{myref}{}
\bibliographystyle{aasjournal}


\listofchanges
\end{document}